\documentclass[preprint,journal]{vgtc}                     

\onlineid{0}

\preprinttext{To appear in IEEE Transactions on Visualization and Computer Graphics.}

\vgtccategory{Research}

\title{Divided Attention Amplifies the Importance of \\Expectation-Aligned Visualization Design}

\author{%
  \authororcid{Jiho Kim}{0000-0002-2289-0927},
  \authororcid{Anna L. Chinni}{0000-0003-2269-734X},
  \authororcid{Karen B. Schloss}{0000-0003-4833-4117},
  \authororcid{Michael Gleicher}{0000-0003-3295-4071}
}

\authorfooter{
  \item
  	Jiho Kim, Anna L. Chinni, Karen B. Schloss, and Michael Gleicher are with University of Wisconsin-Madison.
  	E-mail: {kim999|alchinni|kschloss|gleicher}@wisc.edu.
}

\abstract{Studies have shown that visualization design affects interpretability when visualization interpretation is the user’s sole task. However, in real-world settings, users often engage with visualizations while performing concurrent tasks, such as when users simultaneously monitor alerts or respond to messages. Such divided attention may alter how users interpret visualizations, potentially increasing the importance of designs that align with viewer expectations. We investigated this possibility through two experiments comparing visualization interpretation under single-task and dual-task conditions. Specifically, we examined how well-established inferred mappings between color, spatial position, and semantic concepts affect interpretation when users perform a concurrent task, both with unlimited viewing time (Exp. 1) and under limited viewing time (Exp. 2). Our results show that divided attention amplifies the performance gap between expectation-aligned and expectation-violating designs, affecting response time, interpretation accuracy, and the ability to produce a judgment under time constraints. To explain these results, we model the user’s decision-making process using a Linear Ballistic Accumulator (LBA) framework. Our findings highlight the increased importance of aligning visualization designs with viewer expectations under divided attention and introduce a process-oriented modeling approach to understanding how expectation and multitasking shape visualization interpretation.
}


\teaser{
  \centering
  \includegraphics[width=\linewidth, alt={}]{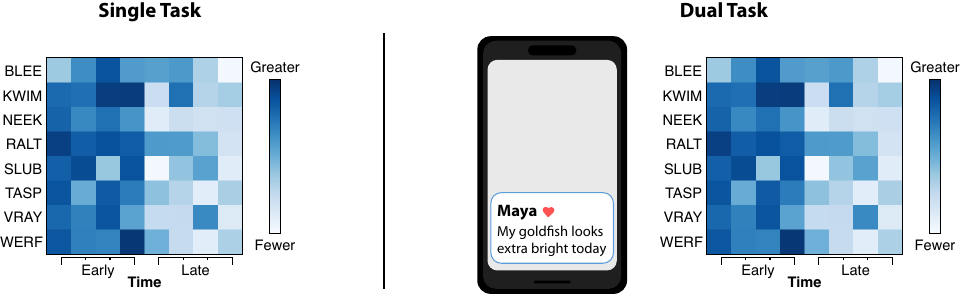}
  \caption{
  	Example trials from the single-task (left) and dual-task (right) conditions in this study. All participants judged which side of a colormap represented greater values by interpreting the legend (colormap task). In the dual-task condition, participants also simultaneously monitored incoming phone messages and ``liked'' messages mentioning pets. Divided attention amplified the performance difference between expectation-aligned and expectation-violating visualization designs in the colormaps task.
  }
  \label{fig:teaser}
}

\graphicspath{{figs/}{figures/}{pictures/}{images/}{./}} 

\usepackage{tabu}                      
\usepackage{booktabs}                  
\usepackage{lipsum}                    
\usepackage{mwe}                       
\usepackage{ccicons}                   

\usepackage{mathptmx}                  

\usepackage{placeins}

\usepackage{dblfloatfix}
\usepackage{cuted}

\usepackage{caption}
\usepackage{float}
\usepackage{pdfpages}
\usepackage{enumitem}

\begin{document}


\firstsection{Introduction}

\maketitle

Most research on how visualization design affects task performance is limited to experiments in which participants focus on a single task, but in many real-world settings, visualization interpretation is only one of several tasks that users must manage simultaneously. To name a few examples, analysts may need to monitor live displays while tracking incoming alerts, students may need to interpret charts displayed during class lectures while managing important text messages from their roommates, and people may need to read maps shown on the news while preparing dinner at home. 
In these situations, other activities compete with visualization interpretation for limited cognitive resources. This competition is an instance of \textit{divided attention}, which includes situations where people must perform concurrent tasks, monitor for intermittent events, respond to interruptions, or rapidly shift attention between information sources \cite{koch2018cognitive}. 

Design choices that appear only mildly costly in single-task settings may become substantially more consequential under conditions of divided attention. For example, Sibrel et al. \cite{sibrel2020relation} proposed that designs which violate observers' prior expectations may be especially difficult to interpret when managing multiple tasks. Focusing on the dark-is-more bias (i.e., expectation that darker colors represent larger magnitudes), they suggested that the small interpretability benefits of encoding larger data values with darker colors (as opposed to lighter colors) observed in single-task experiments could be amplified in more realistic settings with multiple competing tasks. Whether observers extract meaningful patterns in data may hinge on whether the encoded mapping matches their expectations \cite{sibrel2020relation}. 


We tested this possibility in the present study by investigating whether expectation-violating visualization designs become more costly under divided attention. We conducted two experiments using a colormap visualization interpretation task \cite{schloss2019mapping, sibrel2020relation, soto2023more}, in which participants viewed colormap visualizations\footnote{Following \cite{rogowitz1996not, schloss2019mapping, sibrel2020relation, soto2023more}, we use the term `colormap' in reference to visualizations that map gradations of colors on to data magnitude and we use `color scale' in reference to those gradations of color. We note that in the literature, colormaps are also referred to as `heatmaps' and color scales are also referred to as `ramps' or `colormaps'.} (`colormaps' for short) and judged whether greater values were represented on the left or right side (Fig. \ref{fig:teaser}). We varied whether the encoded mapping of the colormap aligned with or violated viewer expectations. Participants completed this as a single-task or in a dual-task condition, in which they also monitored incoming phone messages and made a button press response when certain messages appeared (Fig. \ref{fig:teaser}). 

This study design emulated situations in which viewers interpret visualizations while staying responsive to another stream of information, such as alerts or notifications. In Experiment~1, participants had unlimited time to view and interpret the colormaps, which approximated situations like monitoring static dashboard displays or examining charts in papers. In Experiment~2, participants had a brief, fixed amount of time to view the colormaps, approximating situations like encountering visualizations during a live presentation or seeing them flash on a news channel. 
Across both experiments, we found that divided attention amplified the cost of expectation-violating visualization designs. In Experiment 1 with unlimited viewing time, this effect appeared in measures of response time (RT) and accuracy, whereas in Experiment 2 with limited viewing time, it manifested in miss rate--the likelihood of not being able to produce a response before the trial timed out. 

Although RT, accuracy, and miss rate reveal the behavioral effects of divided attention on interpretations of visualizations, questions remain about the underlying mechanisms. These effects could be due to many possible differences in the decision process. For example, under divided attention, viewers may accumulate evidence more slowly, be more cautious of responding, or begin with a greater bias toward one response. To adjudicate these possibilities, we complemented our analyses of the two experiments with linear ballistic accumulator (LBA) modeling \cite{brown2008simplest}, which models RT, accuracy, and miss rate jointly as the outcome of a single decision process. This allows us to ask not only whether a visualization design is costly, but how it is costly. Using this modeling perspective, we find that divided attention makes the correct interpretation less clearly distinguished from the incorrect one during decision-making, which can explain the amplification of the cost of expectation violation under divided attention.

\textbf{Contributions.} This paper shows that divided attention not only hurts overall performance but also amplifies the cost of visualization designs that conflict with viewer expectations. Specifically, we show:
\begin{itemize}[itemsep=-3pt, topsep=1pt]
    \item Across two experiments, divided attention increases the cost of visualization designs that violate viewers' expectations.
    \item Under time pressure, this amplified cost manifests as failure to produce any judgment in the allotted time.
    \item LBA modeling can explain the amplified cost as a consequence of reduced separation between evidence for correct and incorrect interpretations under divided attention.
\end{itemize}

\section{Related Work}
\subsection{Inferred Mappings on Interpreting Visualizations} \label{sec:inferred_mapping}

When creating data visualizations, designers use perceptual features (e.g., color, shape, size) to encode concepts or values that observers decode to interpret the data. These \textbf{\textit{encoded mappings}} are often specified through legends, labels, or captions, but not always \cite{christen2013colorful}. However, when observers decode the encoded mapping, they do not merely decode legends/labels/captions in a bottom-up fashion. They also use their expectations about the meanings of perceptual features---\textbf{\textit{inferred mappings}}---and have more difficulty interpreting encoded mappings that violate those expectations (see \cite{schloss2025perceptual} for a review).


Several kinds of biases contribute to people's inferred mappings, including the dark-is-more bias \cite{cuff1973colour, mcgranaghan1989ordering, schloss2019mapping, schoenlein2023unifying, sibrel2020relation, soto2023more, bartel2021holey}, opaque-is-more bias \cite{schloss2019mapping, schoenlein2026understanding, bartel2021holey}, saturated-is-more bias \cite{schoenlein2026understanding}, hotspot-is-more bias \cite{sibrel2020relation}, and high-is-more bias \cite{schloss2019mapping, soto2023more, Tversky2011} (see \cite{schloss2025perceptual} for an overview of these biases and how they combine under conflict). Here we focus on the \textbf{\emph{dark-is-more bias} }and the \textbf{\emph{high-is-more bias}} because they are among the most extensively studied and they are the basis for the present study. 

The dark-is-more bias leads to the expectation that darker colors map to larger quantities and the high-is-more bias leads to the expectation that values positioned higher on a legend map to larger quantities. These biases are often studied concurrently using the experimental paradigm we use in the present study (``colormap task'') \cite{schloss2019mapping, sibrel2020relation, soto2023more}, so we explain it here. 
In the initial version of this task, Schloss et al. \cite{schloss2019mapping} presented participants with colormap visualizations representing fictitious data about alien animal sightings over different times of day, along with a legend (Fig. \ref{fig:teaser}). The task was to decode the legend and report whether there were more alien animal sightings early or late in the day by pressing the left/right arrow key. Critically, the encoded mapping in the legend varied across trials, with half of trials mapping darker colors to ``greater'' (dark-more mapping) and half mapping lighter colors to ``greater'' (light-more mapping). Likewise, half of trials had ``greater'' at the top of the legend (high-more mapping) and half had it at the bottom (low-more mapping). Overall, participants were faster at correctly interpreting the colormaps when the legend displayed dark-more mapping (dark-is-more bias) and high-more mapping (high-is-more bias). Moreover, the effect of the dark-is-more bias was stronger under high-more mapping, suggesting that visualizations are especially easy to interpret when designed to align with multiple biases. These results using this paradigm have been replicated and extended in subsequent studies with different kinds of data and different data domains \cite{sibrel2020relation, soto2023more}.

\subsection{Divided Attention and Task Performance}
\label{sec:divided-attention}

Research on effects of visualization design on performance is typically conducted in experiments where participants focus on a single task. In many real-world settings, however, visualization interpretation occurs alongside other activities, requiring viewers to perform two or more concurrent tasks or to ``multitask''. Multitasking usually hurts performance: people tend to become slower, less accurate, or both \cite{awh2000divided,gould1967effects,kahneman1973attention,pashler1994dual}, though evidence suggests training can help people learn to perform certain concurrent tasks well \cite{spelke1976skills, schumacher2001virtually} (see \cite{koch2018cognitive} for a review of dual-task and task switching performance). 

Various theories in cognitive science address constraints in multitasking performance in different ways, but they generally posit that limitations on multitasking reflect some form of shared, limited resource(s) to be applied across tasks (see \cite{musslick2021rationalizing} for a review). Here, we use the term \textbf{\textit{divided attention}} to mean the allocation of limited resource(s) to two or more tasks performed at the same time \cite{kahneman1973attention}.\footnote{We recognize that in many theories concerning limited capacity, the term attention is often used in the context of theories proposing a central limited resource. However, for the purposes of this paper we use it to refer more generally to cognitive resource(s) without making assertions about their nature.} Effects of divided attention are often measured using paradigms that compare performance (e.g., RT, accuracy) of participants who are asked to monitor and respond to two different tasks at the same time (dual-task) to performance of participants on only one task (single-task). The difference in performance serves as a measure of dual-task cost or interference. For example, Strayer and Johnson \cite{strayer2001driven} examined divided attention in a driving simulation and found that participants asked to respond to traffic signals while holding a phone conversation were slower to respond and more likely to miss traffic signals than participants who responded to traffic signals alone. Similarly, Jackson et al. \cite{jackson2023evaluating} found that when participants performed a visual search task along with a word recall task, they missed more search targets and recalled fewer words than if they performed either task alone.

These findings could have important implications for visualization evaluation. Many visualization tasks require viewers to integrate multiple visual elements before responding, rather than simply read off values \cite{bennett1992graphical,pinker1990theory}. For this reason, Bennett and Flach \cite{bennett1992graphical} argued that graphical displays should be evaluated for how well they support divided attention and integrated problem solving. Padilla et al. \cite{padilla2019toward} further suggested that dual-task paradigms and measures such as pupillometry can reveal the cognitive effort required by visualization tasks.

The need to consider attentional constraints is also evident in applied visualization contexts. In heads-up displays, secondary visual information can interfere with concurrent visual tasks \cite{lewis2016through}, although such displays may also improve reaction time in some domain-specific settings \cite{liu2004effects}. Real-time dashboards present a related challenge where users make decisions from partial, local views of dynamic systems \cite{franklin2017dashboard}, motivating interface designs that explicitly monitor and support users' visual attention \cite{toreini2022designing}. These issues arise in monitoring contexts across domains, including factories \cite{maio2024pervasive}, hospitals \cite{franklin2017dashboard}, and cities \cite{stehle2020real,andrienko2015detection}. This body of work suggests that visualization use occurs under attentional constraints, making it important to understand how design choices influence behavior when attention is divided between multiple tasks.


Based on prior work, it would not be surprising if divided attention lowered overall performance on the colormap task in the present study. However, we aimed to test a more nuanced hypothesis, that divided attention would amplify the size of the performance gap between expectation-aligned and expectation-violating designs.

\subsection{Linear Ballistic Accumulator Models}
Studies of human performance often rely on summary measures such as mean RT and accuracy, which are useful for describing behavioral effects, but they conflate potential latent processes. Evidence accumulation models address this limitation by treating observed RT and accuracy patterns as the joint outcome of multiple latent processes, allowing researchers to estimate how much each process contributes to performance \cite{ratcliff1978theory,busemeyer1993decision,usher2001time,ratcliff2008diffusion}. 
The linear ballistic accumulator (LBA) model by Brown and Heathcote \cite{brown2008simplest} models each response option as an accumulator that gathers evidence until one reaches a threshold. LBA can account for response choices and RT distributions for both correct and incorrect responses, while being easier to fit and interpret than other evidence accumulation models such as diffusion-based models \cite{brown2008simplest,donkin2011drawing}. The LBA parameters correspond to distinct components of the decision process. Drift rate ($v$) can be interpreted as the quality of evidence, threshold ($b$) as response caution, starting-point variability ($A$) as the sensitivity to prior bias, and non-decision time ($t_0$) as the motor execution of the final decision. \cite{brown2008simplest,donkin2011drawing,ratcliff2008diffusion}. 

This flexibility explains why LBA models have been adopted across a wide range of domains such as perceptual discriminability, lexical decision, memory, and speed--accuracy tradeoff paradigms \cite{brown2008simplest,donkin2011drawing,ratcliff2008diffusion}. More recently, researchers have also used the model to study biased or value-laden processing. For example, Nishiguchi et al. \cite{nishiguchi2019lba} applied LBA modeling to attentional bias modification and argued that the model revealed cognitive changes that were not recoverable from conventional reaction-time summaries.
At the same time, the literature emphasizes using LBA as a complement to standard behavioral analyses rather than as a replacement. Donkin et al. \cite{donkin2011drawing} frame model-based inference as a process of comparing plausible parameterizations, evaluating model fit, and only then drawing conclusions from parameter estimates. Parameter interpretations are therefore only as convincing as the modeling assumptions and fit diagnostics that support them. 

Returning to visualization research, design effects are often summarized through RT and accuracy, but those aggregate outcomes may reflect multiple subprocesses when viewers interpret visualizations. LBA helps separate them and thus provides a richer account of why some designs are easier or harder to interpret than others.

\section{Experiment 1}
\label{sec:exp-1}

Experiment 1 tested whether divided attention amplifies the cost of colormap designs that violate viewers' expectations when given unlimited viewing time. 
When considering which violations of expectations to study, we focused on the dark-is-more and high-is-more biases for two reasons. First, these biases are well-documented in prior work, providing a basis of comparison when testing for effects of divided attention. Second, an experimental paradigm for studying these biases conjointly already exists \cite{schloss2019mapping} (``colormap task'' in Section \ref{sec:inferred_mapping}), which allowed us to test how divided attention modulates effects of multiple biases (dark-is-more and high-is-more) simultaneously. 

We studied effects of divided attention on the strength of inferred mappings by comparing performance in a single-task condition (colormap task only) and a dual-task condition (colormap task and ``phone message task'') (Fig. \ref{fig:teaser}). In the phone message task, participants read incoming phone messages from fictitious friends and were asked to ``like'' every message that mentioned a pet. We designed the phone message task to create competing attentional demands, requiring participants to consistently monitor incoming messages and split attention between the two tasks to respond. This procedure captured a characteristic of many real-world situations where visualization interpretation occurs alongside other communication and monitoring activities.

Previous work on the dark-is-more and high-is-more biases using this colormap task operationalized these biases in terms of RT (i.e., faster RT when legends specified dark-more and high-more mapping than light-more or low-more mapping, respectively)\cite{schloss2019mapping, sibrel2020relation, soto2023more}. However, we considered that effects of divided attention on this task could manifest in RT, accuracy, or both. Thus, we conducted a pilot study using the same methods as the present experiment to help generate hypotheses. The following preregistered\footnote{Experiment 1 preregistration: https://osf.io/s7h6n} hypotheses are based on prior literature and our pilot data, separated by effects in accuracy and RT. 

\textbf{Accuracy hypotheses.} Overall, the dark-is-more bias will result in fewer errors for dark-more mapping than light-more mapping (\textit{lightness mapping effect}), but this effect will be greater for high-more mapping than low-more mapping (\textit{lightness $\times$ height interaction}). Also overall, divided attention will result in more errors (\textit{divided attention effect}), and will amplify the effects of the dark-is-more bias and high-is-more biases (\textit{task $\times$ lightness mapping interaction} and \textit{task $\times$ height mapping interaction}, respectively). 

\textbf{Response time hypotheses.} Overall, the dark-is-more bias will result in faster RTs for dark-more mapping than for light-more mapping (\textit{lightness mapping effect)}, and the high-is-more bias will result in shorter RTs for high-more than low-more mapping (\textit{height mapping effect)}. Lightness mapping and height mapping will interact, such that the magnitude of the dark-is-more bias will be greater for high-more mapping than low-more mapping (\textit{lightness $\times$ height interaction}) \cite{schloss2019mapping, sibrel2020relation, soto2023more}. Overall, divided attention will result in longer RTs (\textit{divided attention effect}), but based on our pilot data, we did not predict that task would significantly modulate the strength of the dark-is-more or high-is-more biases (\textit{task $\times$ lightness mapping interaction} or \textit{task $\times$ height mapping interaction}, respectively). 

\subsection{Methods}
\label{sec:exp-1-methods}

\textbf{\indent{Participants.}}
Participants were recruited on Prolific \cite{Prolific} in batches of 5 and were eligible if they were at least 18 years old, located in the United States, fluent in English, and had normal or corrected-to-normal vision. We targeted $n=160$ participants, with 80 participants in each task condition, following a power analysis on pilot data using R mixedpower library~\cite{kumle2021estimating} to obtain .80 power for the hypothesized interactions between lightness mapping and task. We recruited 195 participants (single-task: $n$ = 90, dual-task: $n$ = 105) to reach this target after exclusions (see Measures and Exclusions below). Their mean age was 39.8 (range: 19--79), and their reported gender identities included 101 men, 87 women, 4 non-binary, and 3 who preferred not to say. The compensation was \$2.25, with a mean completion time of 9:31 min (median=8:00) and a mean hourly pay of \$14.20 (median=\$16.89). All participants gave informed consent and the UW--Madison IRB approved the protocol.

\textbf{Design, Displays, and Procedure.}
We used a mixed factorial design with three factors: task (single vs.\ dual; between-subject), lightness mapping (dark-more vs.\ light-more; within-subject), and height mapping (high-more vs.\ low-more; within-subject). In the following sections, we first describe the colormap task completed by all participants and then describe the phone message task completed only by participants in the dual-task condition. Participants in the dual-task condition were not instructed to prioritize one task over the other. The detailed instructions for both tasks are in Supplementary Section~\ref{sec:instructions_exp1}.

\textit{\textbf{Colormap task.}} In this task, participants were presented with colormap visualizations representing fictitious data about alien animal sightings at different times of day (\cite{schloss2019mapping}; Section \ref{sec:inferred_mapping}) (Fig.~\ref{fig:teaser}). They reported whether there were more sightings early/late in the day by pressing the left/right arrow key. The participants had unlimited time to respond. Incorrect responses triggered red text displaying ``Incorrect'' for 1000\,ms. We adapted the colormap stimuli from a subset of the colormaps used by Schloss et al. \cite{schloss2019mapping}, designed such that one side of the map had darker values than the other side (Fig.~\ref{fig:teaser}). We tested 20 base colormaps, half with the darker side on the left and half with the darker side on the right. These colormaps used the ColorBrewer Blue color scale \cite{harrower2003colorbrewer} and were displayed on a white background. To the right of the colormap was a vertical color scale, with one end labeled ``greater'' and the other labeled ``fewer.'' In dark-more (D+) mapping, the darker endpoint was labeled ``greater''; in the light-more mapping (L+), the lighter endpoint was labeled ``greater.'' In high-more mapping, ``greater'' appeared at the top of the legend; in the low-more, ``greater'' appeared at the bottom of the legend.
Crossing each base colormap with the two mapping factors yielded 80 experimental stimuli per participant. All 80 stimuli were shown to each participant in randomized order. Before beginning the experiment, participants completed 20 practice trials with stimuli not used in the main experiment. They then completed 80 experimental trials in four blocks of 20, with breaks between blocks. 

\textit{\textbf{Phone message task.}} In the dual-task condition, participants performed the phone message task during the colormap task. A phone panel appeared to the left of the colormap (Fig.~\ref{fig:teaser}), and displayed chat-like messages. Messages appeared one at a time for 3\,s, and the next message appeared after an interval randomly sampled from 0.5--2\,s. Participants pressed the spacebar to ``like'' pet-related messages. A correct ``like'' produced a light-green background, an incorrect ``like'' (false alarm) produced a dark-red background, and missed pet-related messages briefly triggered an angry emoji. 
We generated eighty pet-related messages and 200 non-pet messages using ChatGPT \cite{openai2026chatgpt53}. The non-pet messages were grouped into 20 conversations that shared a theme so that the dialogue flowed naturally. The order of the conversations shown to each participant was randomized. Each message had a 40\% chance to be replaced with a random pet-related message.

\textbf{Measures and Exclusion.}
For the colormap task, we measured trial-level accuracy and RT. We used RTs from correct trials for the RT analysis. We pruned out extreme RTs\footnote{The pruning procedure was unintentionally omitted from the preregistration but it is important because without a response deadline, and extreme RTs may reflect disengagement rather than the decision process of interest. For transparency, we report analyses using the unpruned data in Section~\ref{sec:exp1_rt_unpruned_analysis}.} by log-transforming RT over correct trials to account for distributions with long right tails, and then removed trials with RTs that exceeded $\pm 3$ standard deviations from that participant's mean log-transformed RT. The participant-level distributions and exclusions are shown in Supplementary Section~\ref{sec:exp1_pruned_by_participant}. 

For the phone message task, we recorded counts for hits (liked a pet-related message), misses (failed to like a pet-related message), false-alarms (liked a non-pet message), and correct-rejections (did not like a non-pet message), and computed sensitivity ($d'$) and its 95\% confidence interval (CI) using Gourevitch \& Galanter's (1967) approximation method \cite{gourevitch1967significance}. $d'$ is a measure from signal detection theory that quantifies how well someone can distinguish between signal (e.g., pet-related messages) and noise (non-pet messages), where a value of 0 indicates chance-level performance \cite{tanner1954decision}. 

We preregistered two exclusion criteria based on these measures. For the colormap task, we excluded participants ($n=28$) with $<$50 correct responses out of 80 trials, the minimum accuracy at which the binomial 95\% CI excludes chance performance. For the phone message task, we planned to exclude participants whose 95\% CI for $d'$ included or fell below 0, but none were excluded. The final sample after exclusions was 167 participants (single-task: $n=84$, dual-task: $n=83$).  

\begin{figure}[t]
    \centering
    \includegraphics[width=\linewidth]{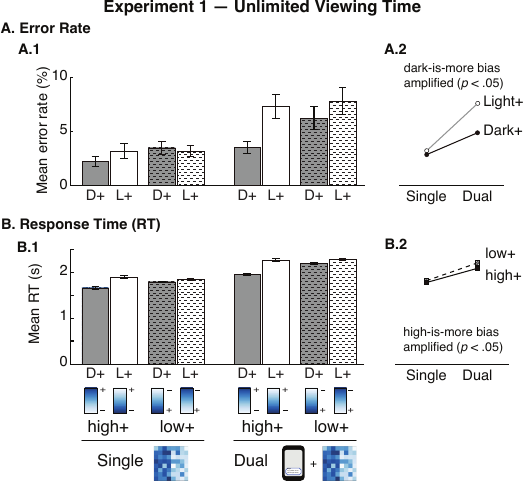}
    \caption{Experiment~1 results. (A.1) Mean error rate and (B.1) mean RT for correct trials grouped by whether greater values were encoded as darker (D+) or lighter (L+), whether the legend specified greater as higher (high+) or lower (low+), and by task condition (single-task or dual-task). 
    (A.2) Mean error rate grouped by lightness mapping and task, showing amplification of the dark-is-more bias (gap between D+ and L+) in the dual-task condition. (B.2) Mean RT grouped by height mapping and task, showing the amplification of the high-is-more bias (gap between high+ and low+) in the dual-task condition.
    Error bars represent standard errors of the means after accounting for overall participant variability (Cousineau-Morey method~\cite{morey2008confidence}). 
    }
    \label{fig:exp1-main}
    \vspace{-6mm}
\end{figure}

\subsection{Results}
Fig.~\ref{fig:exp1-main} shows the mean accuracy in terms of error rate (Fig.~\ref{fig:exp1-main}A) and the mean RT for accurate trials (Fig.~\ref{fig:exp1-main}B), separated by lightness mapping (dark-more/light-more), height mapping (high-more/low-more) and task (single/dual). We analyzed the data using a logistic mixed-effects regression model for accuracy and a linear mixed-effects regression model for RT, with fixed effects for height mapping, lightness mapping, task, and their 2-way and 3-way interactions. In all models, we included by-subject random slopes and intercepts for each within-subject effect. Table~\ref{tab:exp1_model_summary} summarizes key model results and the full model output is in Supplementary Table \ref{tab:exp1_results_acc} (accuracy) and Table \ref{tab:exp1_results_rt_pruned} (RT).

\textbf{Accuracy.} Overall, there was a strong divided attention effect, such that error rate was significantly higher (lower accuracy) in the dual-task than the single-task condition (Figure \ref{fig:exp1-main}A.1, Table~\ref{tab:exp1_model_summary}). Error rate was also significantly higher for low-more than high-more mapping (high-is-more bias), but height mapping did not significantly interact with task, contrary to our hypothesis. Although there was also no overall effect of lightness mapping, a lightness mapping $\times$ task interaction suggested the effect of lightness mapping was greater under divided attention. In Figure \ref{fig:exp1-main}A.1, this difference is apparent in the greater separation between D+ and L+ bars for the dual-task than for the single-task condition. Fig. \ref{fig:exp1-main}A.2 further emphasizes this interaction by showing the effect of lightness mapping as a function of task, averaged over height mapping. Follow-up comparisons within each task condition showed that in the dual-task condition, error rate was lower for dark-more than light-more mapping ($\beta$ = 0.554, $p$ < .001) (dark-is-more bias), but this difference did not reach significance for the single-task condition ($\beta$ = 0.126, $p$ = .461); see Supplementary Table \ref{tab:exp1_results_acc} for full model output. 

Together, these results support the hypothesis that divided attention would amplify the effect of the dark-is-more bias: the dark-is-more bias was too small to have a significant effect on error rate in the single-task condition but the effect was big enough to reach statistical significance in the dual-task condition. Thus, there is greater cost of expectation-violating light-more mappings under divided attention, which increases the likelihood of misinterpreting visualizations. 

\textbf{Response Time.} Overall, there was a strong divided attention effect in RT, with significantly longer RTs overall for the dual-task than the single-task condition (Figure \ref{fig:exp1-main}B.1, Table~\ref{tab:exp1_model_summary}). Replicating prior work \cite{schloss2019mapping, soto2023more, sibrel2020relation}, RTs were significantly faster for high-more than low-more mapping (high-is-more bias) and faster for dark-more than light-more mapping (dark-is-more bias). Also as in prior work, there was a height $\times$ lightness interaction, shown in Fig. \ref{fig:exp1-main}B as the larger difference between the D+ and L+ bars for high-more mapping ($\beta$ = -273.140, $p$ < .001) than for low-more mapping ($\beta$ = -66.508, $p$ = .013) (see Supplementary Table \ref{tab:exp1_results_rt_pruned} for full output of specific comparisons).

Unexpectedly from our pilot data, we also found a significant height $\times$ task interaction, suggesting that the effect of high-is-more bias on RT was amplified under divided attention. Figure \ref{fig:exp1-main}B.2 shows this interaction by plotting the effects of height mapping as a function of task, averaged over lightness mapping. Comparisons within each task condition showed that within the dual-task condition, mean RT was faster for high-more mapping than low-more mapping ($\beta$ = -123.144, $p$ < .001) (high-is-more bias), but this difference was not significant for the single-task condition ($\beta$ = -41.141, $p$ = .155); see Supplementary Table \ref{tab:exp1_results_rt_pruned} for full model output. That is, the high-is-more bias was too small to have a significant effect on RT in the single-task condition but the effect was big enough to reach statistical significance in the dual-task condition.
These results support our hypothesis that divided attention would amplify the effect of the high-is-more bias, but this interaction unexpectedly occurred in RT rather than accuracy. The greater cost of expectation-violating low-more mappings under divided attention increased the time to interpret visualizations correctly. 

\renewcommand{\arraystretch}{0.85}
\begin{table}[t!]
\centering
\small	
\caption{Summary of logistic mixed-effects model (accuracy) and linear mixed-effects model (RT) from Experiment 1. See Supplementary Tables \ref{tab:exp1_results_acc} and \ref{tab:exp1_results_rt_pruned} for full model output.}
\begin{tabular}{l ll ll}
\toprule
& \multicolumn{2}{c}{Accuracy} & \multicolumn{2}{c}{Response Time} \\
\cmidrule(lr){2-3} \cmidrule(lr){4-5}
Predictor & $\beta$ & $p$ & $\beta$ & $p$ \\
\midrule
Intercept   & 3.809 & $<.001$ & 1999.957 & $<.001$ \\
Height & 0.214 & $.036$ & -82.183 & $<.001$\\
Lightness & 0.185 & $.107$ & -170.878  & $<.001$\\
Task   & -0.802 & $<.001$ & 377.518 & $.002$\\
Height:Lightness     & .280 & $.147 $ & -205.738 & $<.001$ \\
Height:Task     & 0.085 & $.678$ & -82.051 & $.031$ \\
Lightness:Task     & 0.460 & $.045$ & -52.867  & $.245$\\
Height:Lightness:Task     & 0.075 & $.846 $ & -52.876 & $.486$\\
\bottomrule
\label{tab:exp1_model_summary}
\vspace{-3em}
\end{tabular}
\end{table}
\renewcommand{\arraystretch}{1}

\textbf{Summary.}
Divided attention amplified the cost of violating viewer expectations. In the dual-task condition, accuracy was lower for light-more than dark-more mappings and responses were slower for low-more than high-more mappings. The cost of expectation violation may thus appear in speed or accuracy, which motivates the process-modeling analysis using LBA reported in Section \ref{sec:LBA}. 

\section{Experiment 2}

In Experiment 1, participants had unlimited time to interpret the colormap visualizations, but in real-world settings, unlimited time is not always available. For example, when attending a lecture, watching the news on TV, or monitoring dynamic visualizations in dashboards, visualizations are displayed for a fixed amount of time, and viewers need to extract information before the visualization changes or disappears.  
Under time pressure, distractions from additional tasks may amplify costs of visualization designs that violate viewer expectations, resulting not only in possible slower or less accurate responses, but also in failures to respond before the visualization is gone \cite{sibrel2020relation}. To investigate effects of divided attention and expectation-violating design under time pressure in Experiment 2, we enforced a limited response window of 1750 ms to impose a meaningful constraint while maintaining task feasibility. Otherwise, the methods were the same as in Experiment 1.

We considered the possibility that effects may manifest differently for Experiment 2 than Experiment 1 because the time constraint could interrupt processing before participants come to a decision, leading them to miss responding before the trials time out. This effect could disproportionately impact expectation-violating conditions that are even harder to interpret under divided attention. 
Thus, we conducted a pilot study to generate hypotheses for Experiment 2 in terms of miss rate, as well as accuracy and RT.
The results from our pilot study inform the following preregistered\footnote{Experiment 2 preregistration: https://osf.io/hux4m} hypotheses.




\textbf{Accuracy hypotheses.} Based on the pilot data, we hypothesized overall effects of lightness mapping (fewer errors for dark-more than light-more mapping) and height mapping (fewer errors for high-more than low-more mapping). We hypothesized no amplification effects due to divided attention because divided attention would slow processing, resulting in more missed trials (see below) rather than more errors from misinterpretations or guessing before trials timed out.  


\textbf{RT and miss rate hypotheses.} 
For RT and miss rate, we predicted a \textit{task} $\times$ \textit{lightness mapping} $\times$ \textit{height mapping interaction}. That is, the interaction between lightness and height mapping (i.e., larger difference between dark-more and light-more mapping under high-more than low-more mapping, also reported in Experiment 1) would be amplified in the dual-task condition compared with the single-task condition. 

If we found this 3-way interaction, we planned follow-up tests for lightness $\times$ height mapping interactions within each task condition, predicting the interaction would reach significance in the dual-task condition, but possibly not in the single-task condition. If this 2-way interaction was significant, we planned comparisons of lightness mapping within each height mapping condition, hypothesizing a significant effect of lightness mapping (dark-is-more bias) under high-more mapping, but the difference might not be significant under low-more mapping. We also planned parallel comparisons and hypotheses for height mapping within each lightness mapping. The logic is that when one mapping aligns with people's expectations, the alignment of the other mapping will matter more. But, when one mapping violates expectations, the additional cost of the second mapping violating expectations is reduced.




\subsection{Methods}

\textbf{\indent{Participants.}}
The participants were recruited on Prolific (same eligibility criteria as Experiment 1, plus participants were ineligible if they did Experiment 1) in batches of 5. We aimed to recruit $n=110$ participants, $n=55$ per task condition, following a power analysis on pilot data to obtain .80 power for the hypothesized three-way interaction. We recruited a total of 124 participants (single-task: $n$ = 59, dual-task: $n$ = 65). Their mean age was 42.2 (range: 19--82), and their reported genders included 65 women, 57 men, 1 non-binary, and 1 who preferred not to say. They were compensated with \$2.00 for this study, which took a mean 9:05 min to complete (median=7:50), resulting in a mean hourly pay of \$13.21 (median=\$15.32). All participants gave informed consent and the UW--Madison IRB approved the protocol.

\textbf{Design, Displays, and Procedure.}
The design, displays, and procedure were the same as Experiment 1 with the following changes to the colormap task. The colormap visualization was displayed for 1750\,ms. If participants did not respond in time, the trial was recorded as missed, a red ``timeout'' message appeared for 500\,ms, and the next trial began after an interval of 500--1000\,ms. If participants responded before timeout, feedback to indicate whether the answer was correct or incorrect was shown for 500\,ms.

\textbf{Measures and Exclusion.}
We measured accuracy, RT, and miss rate (i.e., the proportion of trials in which participants failed to respond within the time limit). 
We excluded participants based on their performance only on the phone task, calculated in the same way we did for the phone task performance in Experiment~1 ($n$ = 7 excluded). We did not prune RTs due to the fixed display duration.
The final analyzed sample was 117 participants (single-task: $n$ = 59, dual-task: $n$ = 58).

\subsection{Results}

\begin{figure*}[t]
    \centering
    \includegraphics[width=.8\linewidth]{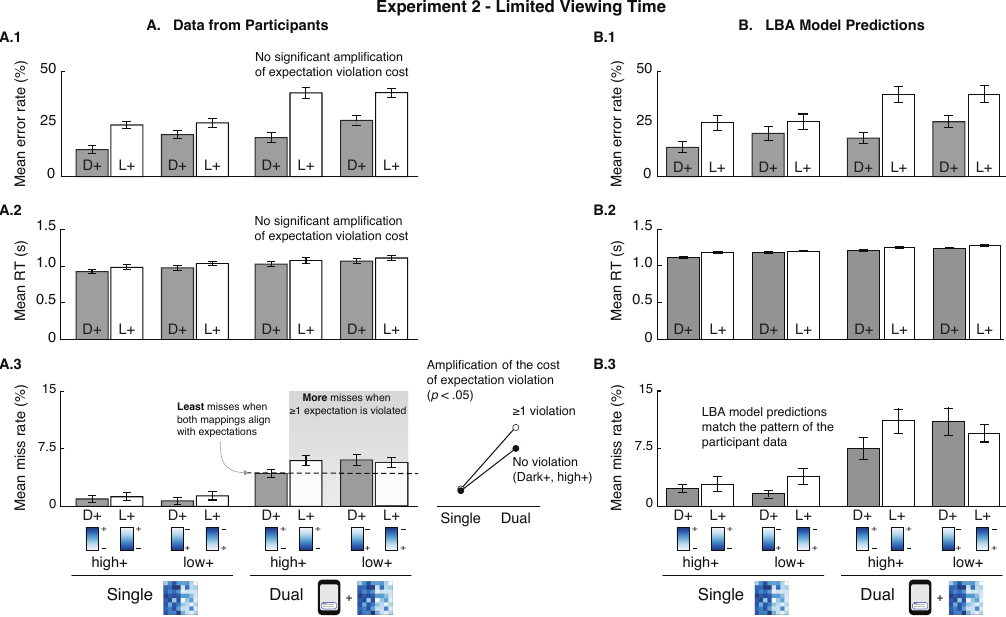}
    \caption{(A) Results of Experiment~2, showing (A.1) mean error rate, (A.2) mean RT, and (A.3) mean miss rate plotted in the same manner as Fig. \ref{fig:exp1-main}. (B) Data simulated from the final LBA model fitted to each Experiment~2 participant for (B.1) mean error rate, (B.2) mean RT, and (B.3) mean miss rate. Error bars represent standard errors of the means after accounting for overall participant variability (Cousineau-Morey method~\cite{morey2008confidence}). Miss rate (bottom) in both real and simulated data show a three-way interaction between task, lightness mapping, and height mapping.}
    \label{fig:exp2-main}
    \vspace{-5.5mm}
\end{figure*}

Fig. \ref{fig:exp2-main}A shows the mean error rate, RT, and miss rate grouped by lightness mapping (dark-more/light-more), height mapping (high-more/low-more) and task (single/dual). We analyzed accuracy (error rate) and RT using the same regression models from Experiment 1, and analyzed miss rate using the same logistic mixed-effects regression model used for accuracy. Table~\ref{tab:exp2_model_summary} summarizes the key model results (see Supplementary Tables \ref{tab:exp2_results_miss}, \ref{tab:exp2_results_rt}, and \ref{tab:exp2_results_acc} for full model output).

\textbf{Accuracy.} As hypothesized, error rates were significantly lower for dark-more than light-more mapping (dark-is-more bias) and for high-more than low-more mapping (high-is-more bias) (Fig. \ref{fig:exp2-main}A, Table \ref{tab:exp2_model_summary}). We also found a significant lightness mapping $\times$ height mapping interaction, in which the effect of lightness mapping was stronger for high-more than low-more mapping (consistent with Experiment 1).
In Fig. \ref{fig:exp2-main}A, this difference is apparent in the greater separation between the D+ and L+ bars for high-more than low-more mapping. Follow-up analyses within each height condition showed that accuracy was higher for dark-more than light-more mapping in both high-more ($\beta$ = 1.167, $p$ < .001) and low-more ($\beta$ = 0.572, $p$ < .001) mappings (see Supplementary Table \ref{tab:exp2_results_acc} for full model output). Thus, among trials completed before timeout, designs that aligned with viewers' expectations led to more accurate interpretation, with an added benefit when both mappings aligned with expectations. 

Unexpectedly based on our pilot data but consistent with Experiment~1, divided attention had an overall effect, with a greater error rate in the dual-task than the single-task condition (Table~\ref{tab:exp2_model_summary}).

\textbf{Response Time.} RT did not show the preregistered three-way interaction, but we did find effects of lightness mapping (faster RTs for dark-more mapping; dark-is-more bias), height mapping (faster RTs for high-more mapping; high-is-more bias), and task (faster RTs for single-task than dual-task; divided attention effect) (Table~\ref{tab:exp2_model_summary}; see Supplementary Table \ref{tab:exp2_results_rt} for full model output).

\textbf{Miss Rate.} For miss rate, we found the hypothesized 3-way interaction, such that the greater difference between dark-more and light-more mapping under high-more mapping was amplified in the dual-task condition (Table \ref{tab:exp2_model_summary}; see Supplementary Table \ref{tab:exp2_results_miss} for full model output and specific comparisons). In Fig. \ref{fig:exp2-main}A.3 this difference is apparent in the greater separation between the D+ and L+ bars under high-more mapping than under low-more mapping in the dual-task condition compared to the single-task condition. 

Follow-up tests for lightness mapping $\times$ height mapping interactions within each task condition showed an interaction in the dual-task condition ($\beta$ = -0.436, $p$ = .034), but not in the single-task condition ($\beta$ = 0.515, $p$ = .216). As shown in Fig. \ref{fig:exp2-main}A.3 (left) and in Supplementary Table \ref{tab:exp2_results_miss}, in the dual-task condition, miss rate was lower for dark-more mapping than light-more mapping under high-more mapping ($\beta$ = -0.423, $p$ = .012), but there was no significant effect of lightness mapping under low-more mapping ($\beta$ = 0.041, $p$ = .771).
The pattern was similar in a parallel analysis within each lightness mapping: the miss rate for high-more mapping was less than for low-more mapping under dark-more mapping ($\beta$ = -0.423, $p$ = .011), but not under light-more mapping ($\beta$ = 0.043, $p$ = .769). 
As shown in the line chart in Fig. \ref{fig:exp2-main}A.3 (right), divided attention in the dual-task condition amplified the effects of $\geq 1$ expectation violations. 
\renewcommand{\arraystretch}{0.85}
\begin{table}[t!]
\centering
\small	
\caption{Summary of logistic mixed-effects models (accuracy and missed trials) and linear mixed-effects model (RT) in Experiment 2. See Supplementary Tables \ref{tab:exp2_results_miss}, \ref{tab:exp2_results_rt}, and \ref{tab:exp2_results_acc} for full model output.}
\begin{tabular}{l ll ll ll}
\toprule
& \multicolumn{2}{c}{Accuracy} & \multicolumn{2}{c}{Response Time} & \multicolumn{2}{c}{Missed Trials}\\
\cmidrule(lr){2-3} \cmidrule(lr){4-5} \cmidrule(lr){6-7}
Predictor & $\beta$ & $p$ & $\beta$ & $p$ & $\beta$ & $p$ \\
\midrule
Intercept           & 1.270 & $<.001$ & 1026.160 & $<.001$  & -3.527 & $<.001$  \\
Height              & 0.315 & $<.001$ & -40.725 & $<.001$   & -0.158 &  $.324$  \\
Lightness           & 0.855 & $<.001$ & -48.054 & $<.001$   & -0.405 & $.010$ \\
Task                & -1.151& $<.001$ & 86.493 & $.037$     &  1.829  & $<.001$ \\
Height:Lightness    & 0.551 & $<.001$ & -21.250  & $.068$   & 0.085 & $.724$ \\
Height:Task         & -0.027& $.854$  & 27.432 & $.086$     & -0.179 & $.489$ \\
Lightness:Task      & 0.269 & $.326$  & -1.464 & $.909$     &  0.300 & $.233$ \\
Height:Light:Task   & 0.108 & $.627$  & -30.957  & $.184$   & -0.963 & $.039$ \\
\bottomrule
\label{tab:exp2_model_summary}
\vspace{-3em}
\end{tabular}
\end{table}
\renewcommand{\arraystretch}{1}

\textbf{Summary.}
When colormaps must be interpreted quickly, divided attention amplified the cost of expectation-violating design. This cost occurred in miss rate, which suggests that mismatching expectations slows processing so observers are less likely to formulate their interpretation in the allotted time when managing multiple tasks. We aim to further understand this process in Section \ref{sec:LBA} on LBA modeling.

\section{LBA Modeling} \label{sec:LBA}
Our behavioral analyses showed that divided attention made expectation-violating designs even more costly. However, RT, accuracy, and miss rate do not by themselves reveal \textit{why} those costs arise; the same behavioral pattern could reflect different cognitive components in the colormap interpretation process. In this section, we complement the behavioral analyses with cognitive modeling to disentangle possible interpretations.

Process modeling is particularly useful because our dependent measures are not independent outcomes. In both experiments, RT and accuracy arise from the same underlying choice process, and in Experiment~2, missed trials add an additional form of failure. A good process model allows us to analyze these measures together and provide a process-oriented explanation of how divided attention and mappings shape visualization interpretation. This approach also provides a basis for predicting performance beyond the exact paradigms studied here.

\begin{figure}[t]
    \centering
    \includegraphics[width=0.9\linewidth]{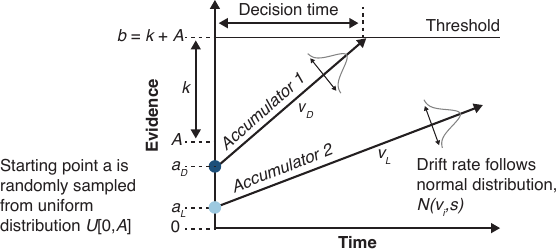}
    \caption{Illustration of the linear ballistic accumulator (LBA) model (figure adapted from Brown and Heathcote \cite{brown2008simplest}). Each response alternative is represented by an accumulator that starts with initial evidence, increases linearly over time, and leads to a response when it reaches threshold. For the colormap task, accumulator 1 represents the darker side, and accumulator 2 represents the lighter side. In this example, $A$ and $b$ are tied across accumulators, but $v$ varies by accumulator.}
    \label{fig:lba-illustration}
    \vspace{-6mm}
\end{figure}

We model the data using a linear ballistic accumulator (LBA) \cite{brown2008simplest,donkin2011drawing}, in which each response option is represented by an accumulator (Fig.~\ref{fig:lba-illustration}). On each trial, the accumulators begin with some amount of starting evidence, accumulate evidence linearly over time, and the first accumulator to reach the threshold determines the response \cite{brown2008simplest}. The main parameters are: starting-point variability ($A$), which captures the sensitivity to pre-decisional bias; response threshold ($b$), which captures the amount of evidence required before responding; drift rate ($v$) and its variance ($s$), which captures how fast evidence is accumulated; and non-decision time ($t_0$), which captures processes outside the decision-making process. In Experiment~2, where responses could time out, this framework allows us to model missed trials as cases in which neither accumulator reaches threshold before the response deadline.

LBA suits our experiments well for three reasons. First, both experiments use a two-choice decision task where each choice can map to an accumulator. Second, LBA can model speed, accuracy, and miss rate jointly through a single decision process. Third, the model affords psychological interpretation while also being able to capture the behavioral patterns of interest in our data \cite{brown2008simplest}.

\subsection{Modeling Strategy and A Priori Assumptions}
Our main goal with LBA was to find a model that could reproduce the key behavioral effects observed in the data, while being simple enough to afford a psychological interpretation of which latent components of the decision process were most affected by divided attention and colormap design. We followed a three-step model selection procedure to achieve this goal. First, we specified a set of theoretically motivated candidate models. Second, we evaluated whether each model could reproduce the core behavioral patterns observed in the data, by fitting the model to each participant and simulating trials. Finally, among models that met this criterion, we selected the best-fitting model using information criteria (AIC/BIC \cite{akaike1974new, schwarz1978estimating}).

The two experiments share the same basic LBA structure. On each trial, the two accumulators, each corresponding to the darker and lighter sides of the colormap, accumulate evidence over time until one of them accumulates enough evidence to reach the threshold and determine the participant's response. 
Experiment~2 added a response deadline which introduced missed trials as an additional outcome. To model missed trials, we computed the probability that neither accumulator reached its respective threshold before trials timed out. 

We narrowed down the possible models using assumptions motivated by the structure of our task and by common LBA practice \cite{donkin2011drawing}. The goal was to keep the model psychologically interpretable by letting the parameters vary only when we had a clear reason to believe that the corresponding psychological process could differ across trials. We note that because task condition (single vs.\ dual) was a between-subject factor and we fit the model separately for each participant, no parameter could be tied across task condition.

\textbf{Starting-point variability ($A$).}
We used $A$ to capture trial-to-trial variability in the initial evidence for each response option. Larger values of $A$ indicate that the starting point for an accumulator varied more across trials, making the decision process more sensitive to the accumulator's initial state. We considered two factors that could plausibly contribute to such variability. First is \textit{accumulator color}; initial evidence may vary depending on whether the accumulator corresponded to the darker or lighter side of the colormap, given the dark-is-more bias. Second is \textit{legend-top color}; initial evidence may vary depending on whether the accumulator corresponded to the color appearing at the top or bottom of the legend, given the high-is-more bias. These assumptions yielded four possible $A$ values per participant. 

\textbf{Response caution ($b$).}
We let $b$ vary across \textit{lightness mapping} and \textit{height mapping} because unexpected mappings can nudge viewers to accumulate more evidence before being confident enough to respond. We also let $b$ vary across \textit{accumulator color}, as viewers might be more hesitant to choose the lighter side than the darker side given the same amount of evidence for both options due to the dark-is-more bias. Thus, $b$ varied across 8 possible conditions within a participant. 

\textbf{Drift rate ($v$).}
We allowed $v$ to vary with \textit{accumulator color}, because the dark-is-more bias may cause evidence to accumulate more quickly for the dark accumulator than the light accumulator. We also let $v$ vary by \textit{lightness mapping} and \textit{height mapping}, because whether a given design aligns with viewers’ expectations can affect how effectively viewers accumulate evidence from the visualization. Thus, $v$ varied across 8 possible conditions within a participant. 


\textbf{Drift rate variance ($s$) and non-decision time ($t_0$).}
We fixed $s=1$ across all participants because the remaining parameters are identifiable when one parameter is fixed. We tied $t_0$ for a participant because we had no reason to expect pressing the left or right arrow key to affect performance after participants had finished interpreting the colormap. 

\begin{figure*}[h]
    \centering
    \includegraphics[width=\linewidth]{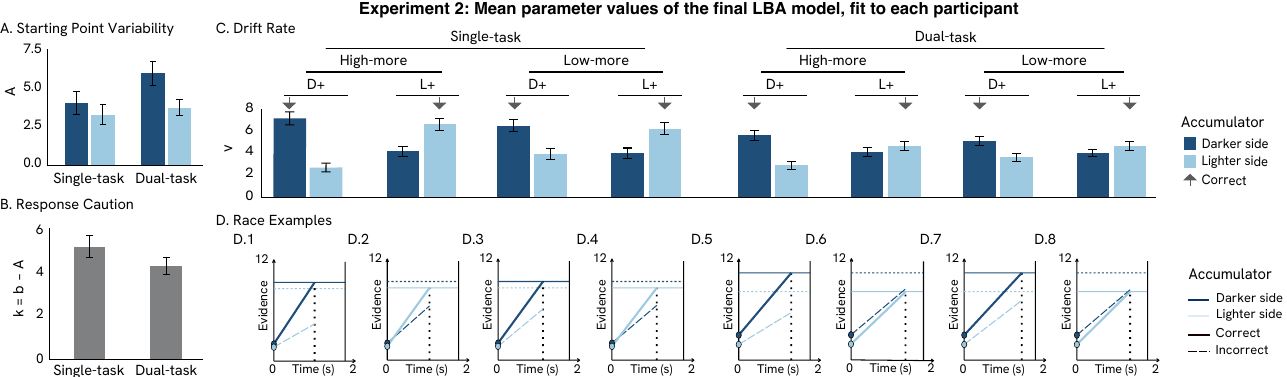}
    \caption{Mean fitted parameter values of the final LBA model for Experiment~2, aggregated across participants. Error bars represent standard error of the means. (A) $A$ varied by accumulator color and task. (B) $k$ varied by task. (C) $v$ varied by accumulator color, task, lightness mapping, and height mapping. The correct accumulator in a given condition is indicated with a gray arrow below x-axis. (D) Each plot depicts an expected scenario of two accumulators in the given experiment condition racing until one of them reaches the response threshold.
    }
    \label{fig:exp2-parameter-means}
    \vspace{-5mm}
\end{figure*}

\subsection{Candidate Simplifications and Final Model Selection}
For each experiment, we began with the most complex model (i.e., free-model) and used the fitted parameter estimates (see Supplementary Fig.~\ref{fig:lba_exp1_free} and Fig.~\ref{fig:lba_exp2_free}) to define simpler candidate models, without sacrificing the model's ability to reproduce the behavioral effects found in the data. 
For $A$, the free-model estimates suggested systematic differences across \textit{accumulator color}, but not across \textit{legend-top color}. Therefore we tied $A$ across \textit{legend-top color}. Larger $A$ values were accompanied by larger $b$ values, so we used $k=b-A$ as a parameter instead of $b$ and tied $k$ across conditions. $k$ has an intuitive psychological interpretation of minimum evidence required before a response \cite{brown2008simplest,donkin2011drawing}. 
Tying $k$ is also reasonable because adjusting response caution separately for each trial is implausible \cite{brown2008simplest,donkin2011drawing}. Last, we retained separate $v$ estimates for each condition. Based on these simplifications, we fit all possible intermediate models. We used L-BFGS-B optimization to fit the model \cite{10.1145/279232.279236} and imposed a sufficiently large upper bound of $A < 20$ to prevent the optimizer from exploring unstable solutions.

We fit each model to each participant and used the models to generate simulated data of 80 trials per participant, matching the number of trials in the experiment. We analyzed the simulated data using the same mixed-effects regression models used in the main analyses. For Experiment 1, the targeted effect to reproduce was the task $\times$ lightness mapping interaction on accuracy and task $\times$ height mapping interaction on RT, as found in the Experiment 1 analysis. For Experiment 2, the targeted effect to reproduce was the task $\times$ lightness mapping $\times$ height mapping interaction on miss rate, as found in the Experiment 2 analysis. The final model for each experiment was selected as the simplest model that reproduced the targeted effects (see Tables~\ref{tab:lba_exp1_candidates_acc},~\ref{tab:lba_exp1_candidates_rt}, and \ref{tab:lba_exp2_candidates} for model evaluations). 

For both experiments within each task condition, the final model allowed $A$ to vary by accumulator color and $v$ to vary by accumulator color, lightness mapping, and height mapping (Table \ref{tab:lba_model_summary}). Fig.~\ref{fig:exp2-parameter-means} shows the mean parameter estimates of the final model for $A$ (Fig.~\ref{fig:exp2-parameter-means}A), $k$ (Fig.~\ref{fig:exp2-parameter-means}B), and $v$ (Fig.~\ref{fig:exp2-parameter-means}C). Using the mean of these parameters, Fig.~\ref{fig:exp2-parameter-means}D shows example accumulator races for the eight experimental conditions, with each plot illustrating how the darker- and lighter-side accumulators increase over time until one reaches its response threshold. Finally, Fig.~\ref{fig:exp2-main}B shows the data simulated with the final Experiment~2 model, which reproduced the original behavioral pattern of Fig.~\ref{fig:exp2-main}A (see Supplementary Fig.~\ref{fig:lba_exp1_final_parameters} for Experiment~1 parameter estimates, and Supplementary Fig.~\ref{fig:lba_exp1_simulated_acc} and~\ref{fig:lba_exp1_simulated_rt} for the data simulated with the final Experiment~1 model).

\renewcommand{\arraystretch}{0.85}
\begin{table}[t!]
\centering
\small	
\caption{Comparison of the final LBA model, the free model (most complex model after a priori assumptions) and null model (simplest model where only $v$ varies by accumulator correctness and all other parameters are tied). \textit{Replicates} refers to whether the model's simulated data predicts the pattern in the original data. Bold text indicates the simplest model that can predict this pattern.}
\begin{tabular}{l lll lll}
\toprule
& \multicolumn{3}{c}{Experiment 1} & \multicolumn{3}{c}{Experiment 2} \\
\cmidrule(lr){2-4} \cmidrule(lr){5-7}
Model & $AIC$ & $BIC$ & $Replicates$ & $AIC$ & $BIC$ & $Replicates$ \\
\midrule
Free model   & 23846 & 29629 & No & 10364 & 16197 & Yes\\
Null model & 23074 & 24451 & No & 15263 & 16616 & No \\
Final model & \textbf{23392} & \textbf{26697} & \textbf{Yes} & \textbf{9791} & \textbf{13125} & \textbf{Yes}\\

\bottomrule
\label{tab:lba_model_summary}
\vspace{-3em}
\end{tabular}
\end{table}
\renewcommand{\arraystretch}{1}

\subsection{Psychological Interpretation of Parameters}
We next interpret the fitted LBA parameters to identify which components of the decision process contributed to the behavioral effects observed in the experiments. We focus on Experiment~2 because the high task accuracy (95.3\%) in Experiment~1 may have restricted the model from identifying accuracy-related decision processes. In contrast, Experiment~2 produced more variation in RT, accuracy, and missed trials, making it more useful in interpreting how LBA parameters relate to these measures. The linear mixed-effects regression analysis of parameters is summarized in Supplementary Table~\ref{tab:lba_exp2_final_parameters}.

\textbf{Expectation violation reduced the separation between correct and incorrect evidence.}
Fig.~\ref{fig:exp2-parameter-means}C shows that the relative drift rates of the darker- and lighter-side accumulators changed with the lightness mapping. In dark-more mappings, the darker-side accumulator has a higher drift rate than the lighter-side accumulator; in light-more mappings, this pattern reversed. This may be because the darker-side accumulator is correct in dark-more mappings, whereas the lighter-side accumulator is correct in light-more mappings. Therefore, we recoded accumulator identity in terms of \textit{correctness} for the remaining analyses. Using this coding, we found that correct accumulators had higher drift rates than incorrect accumulators ($\beta$ = 2.171, $p$ < .001), indicating that evidence accumulated more strongly for the correct interpretation. 

However, this advantage depended on lightness and height mappings. We found a correctness $\times$ lightness mapping interaction ($\beta$ = 1.381, $p$ < .001) and correctness $\times$ height mapping interaction ($\beta$ = 0.824, $p$ < .001).
Follow-up analyses showed that expectation violations reduced the drift-rate separation between the two accumulators. Correct evidence accumulated more slowly in light-more mapping ($\beta$ = 0.576, $p$ < .001) and in low-more mapping ($\beta$ = 0.390, $p$ < .001), whereas incorrect evidence accumulated more quickly in light-more mapping ($\beta$ = -0.805, $p$ < .001) and in low-more mapping ($\beta$ = -0.434, $p$ = .014). This means that both dark-is-more bias and high-is-more bias reduced the separation between correct and incorrect evidence.

The reduced separation can explain Experiment~2 results. A lower correct drift rate made the correct accumulator take longer to reach threshold, increasing RT and also the probability that neither accumulator reached threshold before the deadline (i.e., miss rate). At the same time, a higher incorrect drift rate made the incorrect response more competitive. This increased the probability that the incorrect accumulator won the race, producing lower accuracy. We note that this result differs from a general increase in task difficulty. If expectation-violating designs only slowed processing uniformly, both accumulators would show similar reductions in drift rate. Instead, the fitted parameters show that expectation violation weakened evidence for the correct interpretation while increasing support for the competing interpretation. 

The drift rate also explains the lightness $\times$ height mapping interaction observed for accuracy and miss rate. Overall, there was a 3-way correctness $\times$ lightness mapping $\times$ height mapping interaction ($\beta$ = 1.543, $p$ = .002) in drift rate (Fig. \ref{fig:exp2-parameter-means}C; Supplementary Table~\ref{tab:lba_exp2_final_parameters}). Follow-up analyses showed lightness $\times$ height mapping interactions for both the correct drift rates ($\beta$ = 0.407, $p$ = .016) and incorrect drift rates ($\beta$ = -1.136, $p$ = .001). Expectation violations reduced drift rate separation more when the other mapping aligned with expectations than when it already violated expectations. For example, the effect of lightness mapping on correct and incorrect drift rate was larger in high-more mapping ($\beta$ = 0.779, $p$ < .001 for correct drift rate; $\beta$ = -1.373, $p$ < .001 for incorrect drift rate) than in low-more mapping ($\beta$ = 0.372, $p$ = .004 for correct drift rate; $\beta$ = -0.237, $p$ = .370 for incorrect drift rate). Therefore, evidence separation was most reduced by the first expectation violation, and violating the second expectation produced less additional cost. This pattern parallels the Experiment 2 results on miss rate.

\textbf{Divided attention weakened evidence for the correct interpretation.}
Divided attention affected drift rate, but it did not produce uniform slowing across both accumulators (Fig. \ref{fig:exp2-parameter-means}C; Supplementary Table~\ref{tab:lba_exp2_final_parameters}). A correctness $\times$ task interaction ($\beta$ = -1.565, $p$ < .001) showed that correct-accumulator drift rates were lower in the dual-task condition than in the single-task condition ($\beta$ = -1.616, $p$ = 0.023). In contrast, task did not significantly affect incorrect accumulator drift rates ($\beta$ = -0.052, $p$ = 0.919). This suggests that divided attention weakened evidence supporting the correct interpretation but did not reliably change evidence supporting the incorrect interpretation. As a result, it reduced the separation between the two accumulators.

We did not find evidence that divided attention modulated the effect of expectation violation on drift rate (no task $\times$ lightness mapping ($\beta$ = 0.117, $p$ = .634) or task $\times$ height mapping interaction ($\beta$ = -0.016, $p$ = .947)). However, Fig.~\ref{fig:exp2-parameter-means}D illustrates why a similar reduction in drift-rate gap can have different consequences in single-task and dual-task conditions.
Comparing D.1 and D.2 shows the effect of violating the lightness expectation under single-task/high-more conditions. Although the correct--incorrect drift-rate gap is smaller in D.2 than in D.1, the correct accumulator in D.2 clearly has a steeper trajectory than the incorrect accumulator. Comparing D.5 and D.6 shows the corresponding contrast under dual-task/high-more conditions. Here, the trajectories in D.6 are much closer together, indicating a more competitive race that results in more errors and missed trials. Thus, under dual-task conditions, the correct and incorrect accumulators were already in closer competition and further narrowing the gap made the correct accumulator even less likely to win, explaining the amplified cost of expectation violation.

\textbf{Divided attention increased sensitivity to dark-is-more bias.}
Starting-point variability, $A$, provides evidence for another mechanism. $A$ was higher for dark accumulators than for light accumulators ($\beta$ = 1.488, $p$ < .001; Fig.~\ref{fig:exp2-parameter-means}A, Supplementary Table~\ref{tab:lba_exp2_final_parameters}). A larger $A$ means the accumulator's starting evidence varied across a wider range. This pattern is consistent with the dark-is-more bias favoring the darker side before participants had fully interpreted the legend.

We also found a significant task $\times$ accumulator color interaction for $A$ ($\beta$ = 1.466, $p$ < .001) (Fig.~\ref{fig:exp2-parameter-means}A, Supplementary Table~\ref{tab:lba_exp2_final_parameters}), such that the difference between dark and light accumulators was larger in the dual-task condition than in the single-task condition. 
In the dual-task condition, $A$ was higher for the dark accumulator than for the light accumulator ($\beta$ = 2.221, $p$ = .019), but in the single-task condition this difference was not significant ($\beta$ = 0.755, $p$ = .441).
These results suggest that divided attention increased sensitivity to the initial state of the dark accumulator. On some trials, the dark accumulator could begin closer to threshold before participants had fully decoded the legend. This early advantage would support correct responses in dark-more mapping but could support incorrect responses in light-more mapping. 

This starting-point advantage becomes even more consequential when the drift-rate separation between the correct and incorrect accumulators is small. This effect is illustrated by comparing D.2 and D.6 in Fig.~\ref{fig:exp2-parameter-means}D. Both panels show the light-more/high-more condition, in which the darker-side accumulator corresponds to the incorrect response. In the single-task condition (D.2), the lighter-side accumulator has a sufficiently stronger drift rate that it can overcome the darker accumulator's initial advantage and reach threshold first. In the dual-task condition (D.6), however, the correct and incorrect trajectories are much closer in slope. Under these conditions, variability in the starting point of the darker accumulator becomes more consequential: if the incorrect darker-side accumulator begins with enough initial evidence, it can win the race despite having a slower drift rate than the correct lighter-side accumulator. Thus, under divided attention, weaker drift-rate separation may have increased the importance of initial evidence favoring the darker-side accumulator.



\section{General Discussion}

Across two experiments, we found that divided attention not only lowered performance overall, but also amplified the cost of expectation-violating colormap designs (i.e., misaligning with the dark-is-more and high-is-more biases). In Experiment~1 under unlimited viewing time, the accuracy cost of misaligning with the dark-is-more bias and the RT cost of misaligning with the high-is-more bias were amplified. In Experiment~2 under time pressure, a similar pattern appeared in miss rate, suggesting that the consequence of an expectation-violating design may not only be lower accuracy or slower interpretation, but also a failure to arrive at an interpretation at all. Our findings suggest that expectation-violating designs become especially costly when viewers have more limited attentional resources.


\textbf{Implications for Visualization Design.} 
Our findings imply that evaluations conducted under single-task conditions may underestimate the consequences of expectation-violating designs. Visualization studies should therefore consider the attentional constraints that characterize real use \cite{haroz2012capacity,franklin2017dashboard,toreini2022designing}. Our phone-message task provides one step in this direction by approximating situations where viewers interpret visualizations while managing streams of messages (texts, Slack messages, or emails) from colleagues, friends, or family members. Future work can test the extent to which these effects generalize to other secondary tasks, visualization types, and more realistic work settings.

\textbf{What the LBA Model Adds.}
The LBA analysis helps explain why divided attention did more than just slow interpretation. The model suggests that divided attention reduced the separation between evidence for the correct and incorrect responses. Thus, expectation-violating colormaps became harder to interpret because the available evidence provided less clear support for the correct interpretation. 

The model also suggests that different visualization expectations may influence different stages of decision making. Both the dark-is-more bias and the high-is-more bias affected drift rate, suggesting that these biases influence the interpretation itself. However, the dark-is-more bias also affected the starting-point variability, suggesting an early influence before interpreting the legend. This distinction remains tentative, but it suggests that visualization biases need not share one common cognitive mechanism.

Finally, the LBA analysis shows the value of modeling RT, accuracy, and missed trials as outcomes of one decision process. Behavioral measures can identify when a design causes difficulty, but not whether that difficulty reflects weaker evidence, stronger prior bias, or increased caution. LBA provides one interpretable account of these possibilities. Future work can test this account with other visualization tasks and secondary tasks.

\textbf{Limitations and Open Questions.} This study was an initial step in understanding how divided attention influences effects of expectation alignment on visualization interpretation, but there are several limitations and open questions.  

First, we studied only two kinds of biases (dark-is-more and high-is-more) for one class of visualization task (colormap interpretation). Future work can test the generalizability to other biases,  visualization tasks, and richer visualization settings involving multiple coordinated views or denser dashboards. Moreover, open questions remain concerning the extent to which the amplification effects of divided attention observed here extend beyond inferred mappings to other forms of expectations that influence visualization interpretation, such 
as expectations arising from perceptual organization and salience or expectations about the data itself, such as anticipating particular trends or patterns. 

Second, our dual-task manipulation simulated only one kind of divided attention (i.e., monitoring message streams) in a controlled experimental setting. Questions remain about the effects of divided attention in more realistic settings without extensive repeated trials of the same kind of visualizations. Moreover, in our study, we designed the tasks to minimize the extent to which either task could become automatic---the colormap task required decoding the legend on every trial and the phone message task required reading the messages on every trial to identify pet types not shown in advance. With prior evidence that the cost of multitasking depends on experience and automaticity \cite{spelke1976skills,strobach2017mechanisms}, future work could investigate if the effects reported here change as the primary or secondary task becomes more automated or as participants get better at the tasks through training.

Third, like any cognitive model, an LBA model's conclusions depend on its assumptions and parameterization. Future work could test these assumptions through manipulations designed to target specific decision components. For example, varying response priors could test starting-point effects and changes in display clarity could test drift-rate effects. Such manipulations could provide evidence that the fitted parameters reflect the proposed processes. Another approach would be to collect more direct measures of how participants inspect the visualization, such as eye-tracking data. For example, eye movements could show whether participants spend more time comparing the legend and the colormap, or whether they alternate fixating on each task at different rates depending on the visualization design. Future work could also compare the LBA results with other decision-making models to see whether they lead to the same interpretation.

Finally, the present results suggest an opportunity for design interventions: if expectation-aligned encodings become especially important under divided attention, then attention-aware or context-aware visualization systems may be able to adapt their presentation when rapid interpretation is most critical.

\textbf{Conclusion.} This work shows that divided attention can amplify the cost of visualization designs that violate viewers' expectations. Expectation-violating lightness and height mappings became more costly under divided attention, and under time pressure these costs also appeared as failures to produce a judgment within the allotted time. The LBA analysis suggests that these costs reflected changes in the decision process, especially reduced separation between evidence for correct and incorrect interpretations. These findings suggest that aligning visual encodings with viewer expectations becomes especially important when viewers are likely to be interrupted or distracted, and highlight the value of evaluating visualization designs under attentional constraints that better reflect many real-world uses of visualizations.


\section*{Acknowledgments}
We thank Ami Eidels for helpful feedback on modeling and C. Shawn Green for guidance on statistical analyses. This work was supported by the National Science Foundation under Grant No.~2007436, Grant No.~2147044, and Grant No.~2419493.

\section*{Supplemental Materials}
The Supplementary Material includes experiment procedures, instructions, datasets, analysis code, and results of all analyses in Experiments 1 and 2. It also includes the code used to fit and analyze LBA models, evaluations of all candidate models, and parameter analyses of the final model. 

\section*{AI Use Statement}
GPT-5.3 and GPT-5.4 were used to assist with generating experimental stimuli, writing analysis code, producing figures, and editing text for clarity and grammar. All outputs were reviewed by the authors.

\vspace{-0.5em}

\bibliographystyle{abbrv-doi-hyperref}
\bibliography{reference}

@article{akaike1974new,
  title={A new look at the statistical model identification},
  author={Akaike, Hirotugu},
  journal={IEEE transactions on automatic control},
  volume={19},
  number={6},
  pages={716--723},
  year={1974},
  publisher={Ieee},
  doi = {10.1109/TAC.1974.1100705}
}

@article{tanner1954decision,
  title={A decision-making theory of visual detection.},
  author={Tanner Jr, Wilson P and Swets, John A},
  journal={Psychological review},
  volume={61},
  number={6},
  pages={401},
  year={1954},
  publisher={American Psychological Association}
}

@inproceedings{andrienko2015detection,
  author={Andrienko, Natalia and Andrienko, Gennady and Fuchs, Georg and Rinzivillo, Salvatore and Betz, Hans-Dieter},
  booktitle={2015 IEEE International Conference on Data Science and Advanced Analytics (DSAA)}, 
  title={Detection, tracking, and visualization of spatial event clusters for real time monitoring}, 
  year={2015},
  volume={},
  number={},
  pages={1-10},
  doi={10.1109/DSAA.2015.7344880}
}

@article{awh2000divided,
  title   = {Divided attention and visual search for simple versus complex features},
  author  = {Awh, Edward and Pashler, Harold},
  journal = {Perception \& Psychophysics},
  volume  = {62},
  number  = {5},
  pages   = {970--982},
  year    = {2000},
  doi = {10.1016/S0042-6989(03)00339-0}
}

@article{bartel2021holey,
  title={A holey perspective on venn diagrams},
  author={Bartel, Anna N and Lande, Kevin J and Roos, Joris and Schloss, Karen B},
  journal={Cognitive Science},
  volume={46},
  number={1},
  pages={e13073},
  year={2021},
  publisher={Wiley Online Library}, 
  doi = {10.1111/cogs.13073}
}

@article{bennett1992graphical,
  title     = {Graphical displays: Implications for divided attention, focused attention, and problem solving},
  author    = {Bennett, Kevin B and Flach, John M},
  journal   = {Human factors},
  volume    = {34},
  number    = {5},
  pages     = {513--533},
  year      = {1992},
  publisher = {SAGE Publications Sage CA: Los Angeles, CA},
  doi = {10.1177/001872089203400502}
}

@article{brown2008simplest,
  title   = {The simplest complete model of choice response time: Linear ballistic accumulation},
  author  = {Brown, Scott D. and Heathcote, Andrew},
  journal = {Cognitive Psychology},
  volume  = {57},
  number  = {3},
  pages   = {153--178},
  year    = {2008},
  doi     = {10.1016/j.cogpsych.2007.12.002}
}

@article{busemeyer1993decision,
  title   = {Decision field theory: A dynamic-cognitive approach to decision making in an uncertain environment},
  author  = {Busemeyer, Jerome R. and Townsend, James T.},
  journal = {Psychological Review},
  volume  = {100},
  number  = {3},
  pages   = {432--459},
  year    = {1993},
  doi     = {10.1037/0033-295X.100.3.432}
}

@article{christen2013colorful,
  title     = {Colorful brains: 14 years of display practice in functional neuroimaging},
  author    = {Christen, Markus and Vitacco, Deborah A and Huber, Lara and Harboe, Julie and Fabrikant, Sara I and Brugger, Peter},
  journal   = {NeuroImage},
  volume    = {73},
  pages     = {30--39},
  year      = {2013},
  publisher = {Elsevier},
  doi = {10.1016/j.neuroimage.2013.01.068}
}

@article{cuff1973colour,
  title     = {Colour on temperature maps},
  author    = {Cuff, David J},
  journal   = {The Cartographic Journal},
  volume    = {10},
  number    = {1},
  pages     = {17--21},
  year      = {1973},
  publisher = {Taylor \& Francis},
  doi = {10.1179/caj.1973.10.1.17}
}

@article{donkin2011drawing,
  title   = {Drawing conclusions from choice response time models: A tutorial using the linear ballistic accumulator},
  author  = {Donkin, Chris and Brown, Scott and Heathcote, Andrew},
  journal = {Journal of Mathematical Psychology},
  volume  = {55},
  number  = {2},
  pages   = {140--151},
  year    = {2011},
  doi     = {10.1016/j.jmp.2010.10.001}
}

@article{franklin2017dashboard,
  title     = {Dashboard visualizations: Supporting real-time throughput decision-making},
  author    = {Franklin, Amy and Gantela, Swaroop and Shifarraw, Salsawit and Johnson, Todd R and Robinson, David J and King, Brent R and Mehta, Amit M and Maddow, Charles L and Hoot, Nathan R and Nguyen, Vickie and others},
  journal   = {Journal of biomedical informatics},
  volume    = {71},
  pages     = {211--221},
  year      = {2017},
  publisher = {Elsevier},
  doi       = {10.1016/j.jbi.2017.05.024}
}

@article{gould1967effects,
  title   = {The effects of divided attention on visual monitoring of multi-channel displays},
  author  = {Gould, John D. and Schaffer, Amy},
  journal = {Human Factors},
  volume  = {9},
  number  = {3},
  pages   = {191--202},
  year    = {1967},
  doi = {10.1177/001872086700900301}
}

@article{gourevitch1967significance,
  title={A significance test for one parameter isosensitivity functions},
  author={Gourevitch, Vivian and Galanter, Eugene},
  journal={Psychometrika},
  volume={32},
  number={1},
  pages={25--33},
  year={1967},
  publisher={Springer-Verlag},
  doi = {10.1007/BF02289402}
}

@article{haroz2012capacity,
  title   = {How capacity limits of attention influence information visualization effectiveness},
  author  = {Haroz, Steve and Whitney, David},
  journal = {IEEE Transactions on Visualization and Computer Graphics},
  volume  = {18},
  number  = {12},
  pages   = {2402--2410},
  year    = {2012},
  doi     = {10.1109/TVCG.2012.233}
}

@article{harrower2003colorbrewer,
  title={ColorBrewer. org: an online tool for selecting colour schemes for maps},
  author={Harrower, Mark and Brewer, Cynthia A},
  journal={The Cartographic Journal},
  volume={40},
  number={1},
  pages={27--37},
  year={2003},
  publisher={Taylor \& Francis},
  doi = {10.1179/000870403235002042}
}

@article{jackson2023evaluating,
  title={Evaluating the dual-task decrement within a simulated environment: Word recall and visual search},
  author={Jackson, Kenneth M and Shaw, Tyler H and Helton, William S},
  journal={Applied Ergonomics},
  volume={106},
  pages={103861},
  year={2023},
  publisher={Elsevier},
  doi = {10.1016/j.apergo.2022.103861}
}

@book{kahneman1973attention,
	author = {Daniel Kahneman},
	publisher = {Prentice-Hall},
	title = {Attention and Effort},
	year = {1973}
}

@article{koch2018cognitive,
  title={Cognitive structure, flexibility, and plasticity in human multitasking—An integrative review of dual-task and task-switching research.},
  author={Koch, Iring and Poljac, Edita and M{\"u}ller, Hermann and Kiesel, Andrea},
  journal={Psychological bulletin},
  volume={144},
  number={6},
  pages={557},
  year={2018},
  publisher={American Psychological Association},
  doi = {10.1037/bul0000144}
}

@article{kumle2021estimating,
  title={Estimating power in (generalized) linear mixed models: An open introduction and tutorial in R},
  author={Kumle, Levi and V{\~o}, Melissa L-H and Draschkow, Dejan},
  journal={Behavior research methods},
  volume={53},
  number={6},
  pages={2528--2543},
  year={2021},
  publisher={Springer},
  doi = {10.3758/s13428-021-01546-0}
}

@article{lewis2016through,
  title     = {Through the Google Glass: The impact of heads-up displays on visual attention},
  author    = {Lewis, Joanna E and Neider, Mark B},
  journal   = {Cognitive research: principles and implications},
  volume    = {1},
  number    = {1},
  pages     = {13},
  year      = {2016},
  publisher = {Springer},
  doi       = {10.1186/s41235-016-0015-6}
}

@article{liu2004effects,
  title   = {Effects of using head-up display in automobile context on attention demand and driving performance},
  author  = {Liu, Yung-Ching},
  journal = {Displays},
  volume  = {25},
  number  = {4},
  pages   = {157--165},
  year    = {2004},
  doi     = {10.1016/j.displa.2004.01.001}
}

@article{maio2024pervasive,
  title     = {Pervasive Augmented Reality to support real-time data monitoring in industrial scenarios: Shop floor visualization evaluation and user study},
  author    = {Maio, Rafael and Ara{\'u}jo, Tiago and Marques, Bernardo and Santos, Andr{\'e} and Ramalho, Pedro and Almeida, Duarte and Dias, Paulo and Santos, Beatriz Sousa},
  journal   = {Computers \& Graphics},
  volume    = {118},
  pages     = {11--22},
  year      = {2024},
  publisher = {Elsevier},
  doi = {10.1016/j.cag.2023.10.025}
}

@article{mcgranaghan1989ordering,
  title     = {Ordering choropleth map symbols: The effect of background},
  author    = {McGranaghan, Matthew},
  journal   = {The American Cartographer},
  volume    = {16},
  number    = {4},
  pages     = {279--285},
  year      = {1989},
  publisher = {Taylor \& Francis},
  doi = {10.1559/152304089783813918}
}

@article{morey2008confidence,
    author = {Morey, Richard},
    year = {2008},
    month = {09},
    pages = {61-64},
    title = {Confidence Intervals from Normalized Data: A correction to Cousineau (2005)},
    volume = {4},
    journal = {Tutorials in Quantitative Methods for Psychology},
    doi = {10.20982/tqmp.04.2.p061}
}

@article{musslick2021rationalizing,
  title={Rationalizing constraints on the capacity for cognitive control},
  author={Musslick, Sebastian and Cohen, Jonathan D},
  journal={Trends in cognitive sciences},
  volume={25},
  number={9},
  pages={757--775},
  year={2021},
  publisher={Elsevier},
  doi = {10.1016/j.tics.2021.06.001}
}

@article{nishiguchi2019lba,
  title   = {Linear ballistic accumulator modeling of attentional bias modification revealed disturbed evidence accumulation of negative information by explicit instruction},
  author  = {Nishiguchi, Yuki and Sakamoto, Jiro and Kunisato, Yoshihiko and Takano, Keisuke},
  journal = {Frontiers in Psychology},
  volume  = {10},
  pages   = {2447},
  year    = {2019},
  doi     = {10.3389/fpsyg.2019.02447}
}

@misc{openai2026chatgpt53,
  author       = {OpenAI},
  title        = {ChatGPT (GPT-5.3 Instant)},
  year         = {2026},
  howpublished = {Large language model},
  url          = {https://chat.openai.com}
}

@ARTICLE{padilla2019toward,
  author={Padilla, Lace M.K. and Castro, Spencer C. and Quinan, P. Samuel and Ruginski, Ian T. and Creem-Regehr, Sarah H.},
  journal={IEEE Transactions on Visualization and Computer Graphics}, 
  title={Toward Objective Evaluation of Working Memory in Visualizations: A Case Study Using Pupillometry and a Dual-Task Paradigm}, 
  year={2020},
  volume={26},
  number={1},
  pages={332-342},
  doi={10.1109/TVCG.2019.2934286}}

@article{pashler1994dual,
  title     = {Dual-task interference in simple tasks: data and theory.},
  author    = {Pashler, Harold},
  journal   = {Psychological bulletin},
  volume    = {116},
  number    = {2},
  pages     = {220},
  year      = {1994},
  publisher = {American Psychological Association},
  doi = {10.1037/0033-2909.116.2.220}
}

@article{pinker1990theory,
  title={A theory of graph comprehension},
  author={Pinker, Steven},
  journal={Artificial intelligence and the future of testing},
  volume={73},
  number={126},
  pages={1--2},
  year={1990},
  publisher={Hillsdale, NJ}
}

@misc{Prolific,
  author       = {{Prolific}},
  title        = {Prolific},
  howpublished = {Online participant recruitment platform},
  year         = {2026},
  url          = {https://www.prolific.com}
}

@article{ratcliff1978theory,
  title   = {A theory of memory retrieval},
  author  = {Ratcliff, Roger},
  journal = {Psychological Review},
  volume  = {85},
  number  = {2},
  pages   = {59--108},
  year    = {1978},
  doi     = {10.1037/0033-295X.85.2.59}
}

@article{ratcliff2008diffusion,
  title   = {The diffusion decision model: Theory and data for two-choice decision tasks},
  author  = {Ratcliff, Roger and McKoon, Gail},
  journal = {Neural Computation},
  volume  = {20},
  number  = {4},
  pages   = {873--922},
  year    = {2008},
  doi     = {10.1162/neco.2008.12-06-420}
}

@article{rogowitz1996not,
author = {Rogowitz, Bernice E. and Treinish, Lloyd A. and Bryson, Steve},
title = {How not to lie with visualization},
year = {1996},
issue_date = {May/June 1996},
publisher = {American Institute of Physics Inc.},
address = {USA},
volume = {10},
number = {3},
issn = {0894-1866},
url = {https://doi.org/10.1063/1.4822401},
doi = {10.1063/1.4822401},
journal = {Comput. Phys.},
month = jun,
pages = {268–273},
numpages = {6}
}

@article{schloss2025perceptual,
  title={Perceptual and cognitive foundations of information visualization},
  author={Schloss, Karen B},
  journal={Annual Review of Vision Science},
  volume={11},
  number={1},
  pages={303--330},
  year={2025},
  publisher={Annual Reviews},
  doi = {10.1146/annurev-vision-110323-110009}
}

@article{schloss2019mapping,
  title     = {Mapping color to meaning in colormap data visualizations},
  author    = {Schloss, Karen B and Gramazio, Connor C and Silverman, Allison T and Parker, Madeline L and Wang, Audrey S},
  journal   = {IEEE transactions on visualization and computer graphics},
  volume    = {25},
  number    = {1},
  pages     = {810--819},
  year      = {2019},
  publisher = {IEEE},
  doi       = {10.1109/TVCG.2018.2865147}
}

@ARTICLE{schoenlein2023unifying,
  author={Schoenlein, Melissa A. and Campos, Johnny and Lande, Kevin J. and Lessard, Laurent and Schloss, Karen B.},
  journal={IEEE Transactions on Visualization and Computer Graphics}, 
  title={Unifying Effects of Direct and Relational Associations for Visual Communication}, 
  year={2023},
  volume={29},
  number={1},
  pages={385-395},
  doi={10.1109/TVCG.2022.3209443}}

@article{schoenlein2026understanding,
  title={Understanding the opaque-is-more bias and saturated-is-more bias for colormap data visualizations},
  author={Schoenlein, Melissa A and Sidibe, Mouloukou and Schloss, Karen B},
  journal={Attention, Perception, \& Psychophysics},
  volume={88},
  number={3},
  pages={69},
  year={2026},
  publisher={Springer}
}

@article{schumacher2001virtually,
  title={Virtually perfect time sharing in dual-task performance: Uncorking the central cognitive bottleneck},
  author={Schumacher, Eric H and Seymour, Travis L and Glass, Jennifer M and Fencsik, David E and Lauber, Erick J and Kieras, David E and Meyer, David E},
  journal={Psychological science},
  volume={12},
  number={2},
  pages={101--108},
  year={2001},
  publisher={SAGE Publications Sage CA: Los Angeles, CA},
  doi = {10.1111/1467-9280.00318}
}

@article{schwarz1978estimating,
  title={Estimating the dimension of a model},
  author={Schwarz, Gideon},
  journal={The annals of statistics},
  pages={461--464},
  year={1978},
  publisher={JSTOR},
  doi = {10.1214/aos/1176344136}
}

@article{sibrel2020relation,
  title     = {The relation between color and spatial structure for interpreting colormap data visualizations},
  author    = {Sibrel, Shannon C and Rathore, Ragini and Lessard, Laurent and Schloss, Karen B},
  journal   = {Journal of vision},
  volume    = {20},
  number    = {12},
  pages     = {7--7},
  year      = {2020},
  publisher = {The Association for Research in Vision and Ophthalmology},
  doi = {10.1167/jov.20.12.7}
}

@article{soto2023more,
  title     = {More of what? Dissociating effects of conceptual and numeric mappings on interpreting colormap data visualizations},
  author    = {Soto, Alexis and Schoenlein, Melissa A and Schloss, Karen B},
  journal   = {Cognitive Research: Principles and Implications},
  volume    = {8},
  number    = {1},
  pages     = {38},
  year      = {2023},
  publisher = {Springer},
  doi = {10.1186/s41235-023-00482-1}
}

@article{spelke1976skills,
  title={Skills of divided attention},
  author={Spelke, Elizabeth and Hirst, William and Neisser, Ulric},
  journal={Cognition},
  volume={4},
  number={3},
  pages={215--230},
  year={1976},
  publisher={Elsevier},
  doi = {10.1016/0010-0277(76)90018-4}
}

@article{stehle2020real,
  title     = {Real-time and archival data visualisation techniques in city dashboards},
  author    = {Stehle, Samuel and Kitchin, Rob},
  journal   = {International Journal of Geographical Information Science},
  volume    = {34},
  number    = {2},
  pages     = {344--366},
  year      = {2020},
  publisher = {Taylor \& Francis},
  doi = {10.1080/13658816.2019.1594823}
}

@article{strayer2001driven,
  title={Driven to distraction: Dual-task studies of simulated driving and conversing on a cellular telephone},
  author={Strayer, David L and Johnston, William A},
  journal={Psychological science},
  volume={12},
  number={6},
  pages={462--466},
  year={2001},
  publisher={Sage Publications Sage CA: Los Angeles, CA},
  doi = {10.1111/1467-9280.00386}
}

@article{strobach2017mechanisms,
  title={Mechanisms of practice-related reductions of dual-task interference with simple tasks: data and theory},
  author={Strobach, Tilo and Torsten, Schubert},
  journal={Advances in cognitive psychology},
  volume={13},
  number={1},
  pages={28},
  year={2017},
  doi = {10.5709/acp-0204-7}
}

@article{toreini2022designing,
  title   = {Designing attentive information dashboards},
  author  = {Toreini, Peyman and Langner, Moritz and Maedche, Alexander and Morana, Stefan and Vogel, Tobias},
  journal = {Journal of the Association for Information Systems},
  volume  = {23},
  number  = {2},
  pages   = {521--552},
  year    = {2022},
  doi     = {10.17705/1jais.00732}
}

@article{Tversky2011,
  author   = {Tversky, Barbara},
  title    = {Visualizing Thought},
  journal  = {Topics in Cognitive Science},
  volume   = {3},
  number   = {3},
  pages    = {499-535},
  doi      = {10.1111/j.1756-8765.2010.01113.x},
  url      = {https://onlinelibrary.wiley.com/doi/abs/10.1111/j.1756-8765.2010.01113.x},
  eprint   = {https://onlinelibrary.wiley.com/doi/pdf/10.1111/j.1756-8765.2010.01113.x},
  year     = {2011}
}

@article{usher2001time,
  title   = {The time course of perceptual choice: The leaky, competing accumulator model},
  author  = {Usher, Marius and McClelland, James L.},
  journal = {Psychological Review},
  volume  = {108},
  number  = {3},
  pages   = {550--592},
  year    = {2001},
  doi     = {10.1037/0033-295X.108.3.550}
}

@article{10.1145/279232.279236,
author = {Zhu, Ciyou and Byrd, Richard H. and Lu, Peihuang and Nocedal, Jorge},
title = {Algorithm 778: L-BFGS-B: Fortran subroutines for large-scale bound-constrained optimization},
year = {1997},
issue_date = {Dec. 1997},
publisher = {Association for Computing Machinery},
address = {New York, NY, USA},
volume = {23},
number = {4},
issn = {0098-3500},
url = {https://doi.org/10.1145/279232.279236},
doi = {10.1145/279232.279236},
journal = {ACM Trans. Math. Softw.},
month = dec,
pages = {550–560},
numpages = {11}
}

\clearpage
\appendix 
\onecolumn

\renewcommand*{\thesection}{S}
\counterwithin{figure}{section}
\counterwithin{table}{section}


\section{Supplementary Material}\label{sec:supplementary}
\raggedbottom

\subsection{Experiment Procedure}
\begin{figure}[!htbp]
    \centering
    \includegraphics[width=\linewidth]{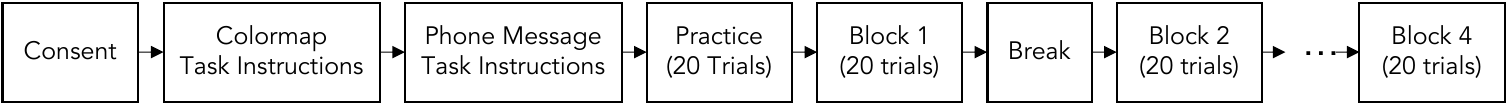}
    \caption{The flow of both experiment 1 and 2. The ``Phone Message Task Instructions'' were only displayed to participants in the dual-task condition. After each block (after block 1, 2, and 3), there was a break where participants could take as much time as the want before proceeding to the next block.}
    \label{fig:exp_flow}
\end{figure}

\FloatBarrier

\subsection{Experiment 1 Instructions}
\label{sec:instructions_exp1}

\begin{figure}[H]
    \centering
    \includegraphics[width=\textwidth]{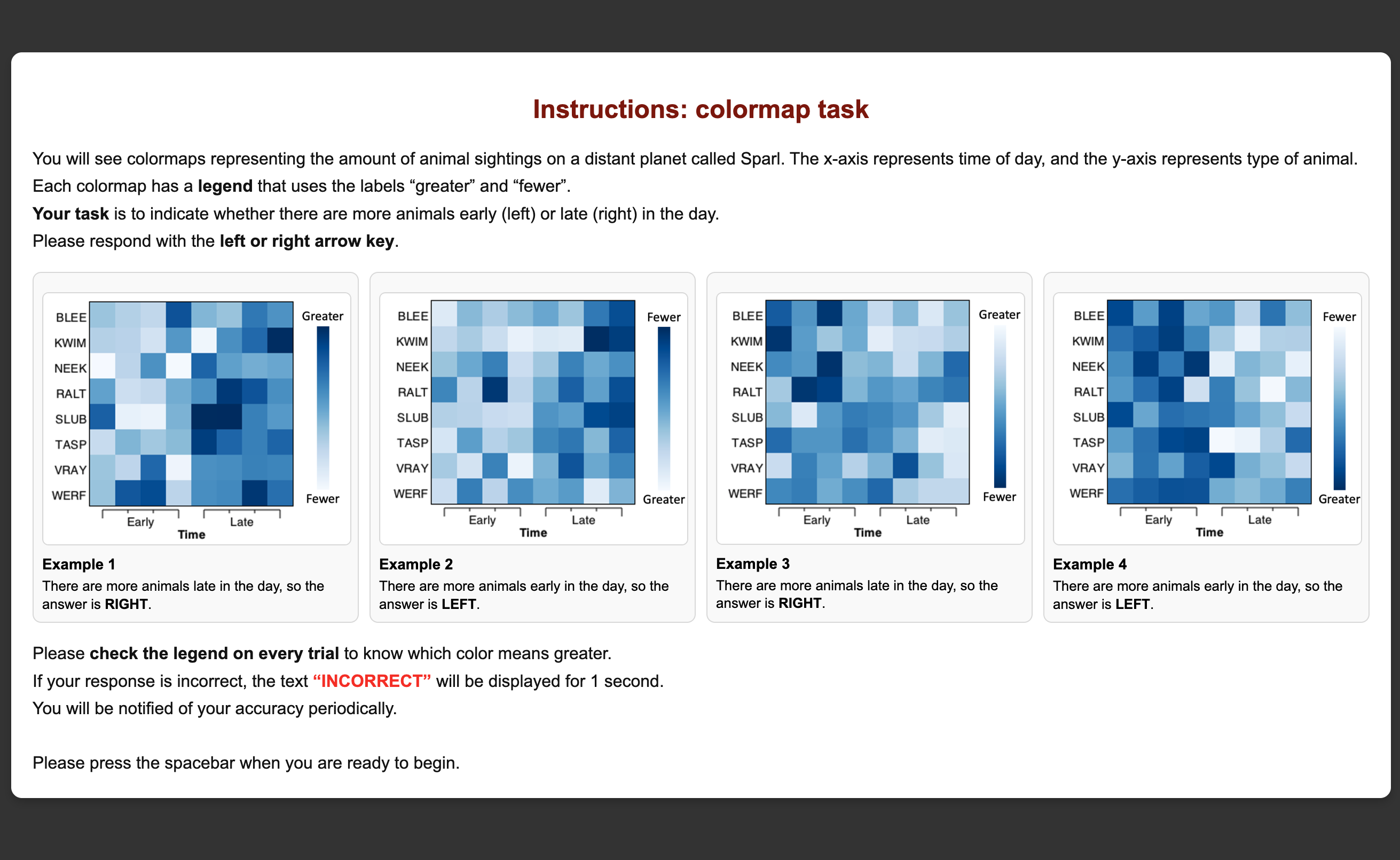}
    \caption{Experiment 1 colormap task instructions for the participants in the single-task condition.}
    \label{fig:instruction-colormap-single}
\end{figure}

\begin{figure}[H]
    \centering
    \includegraphics[width=\textwidth]{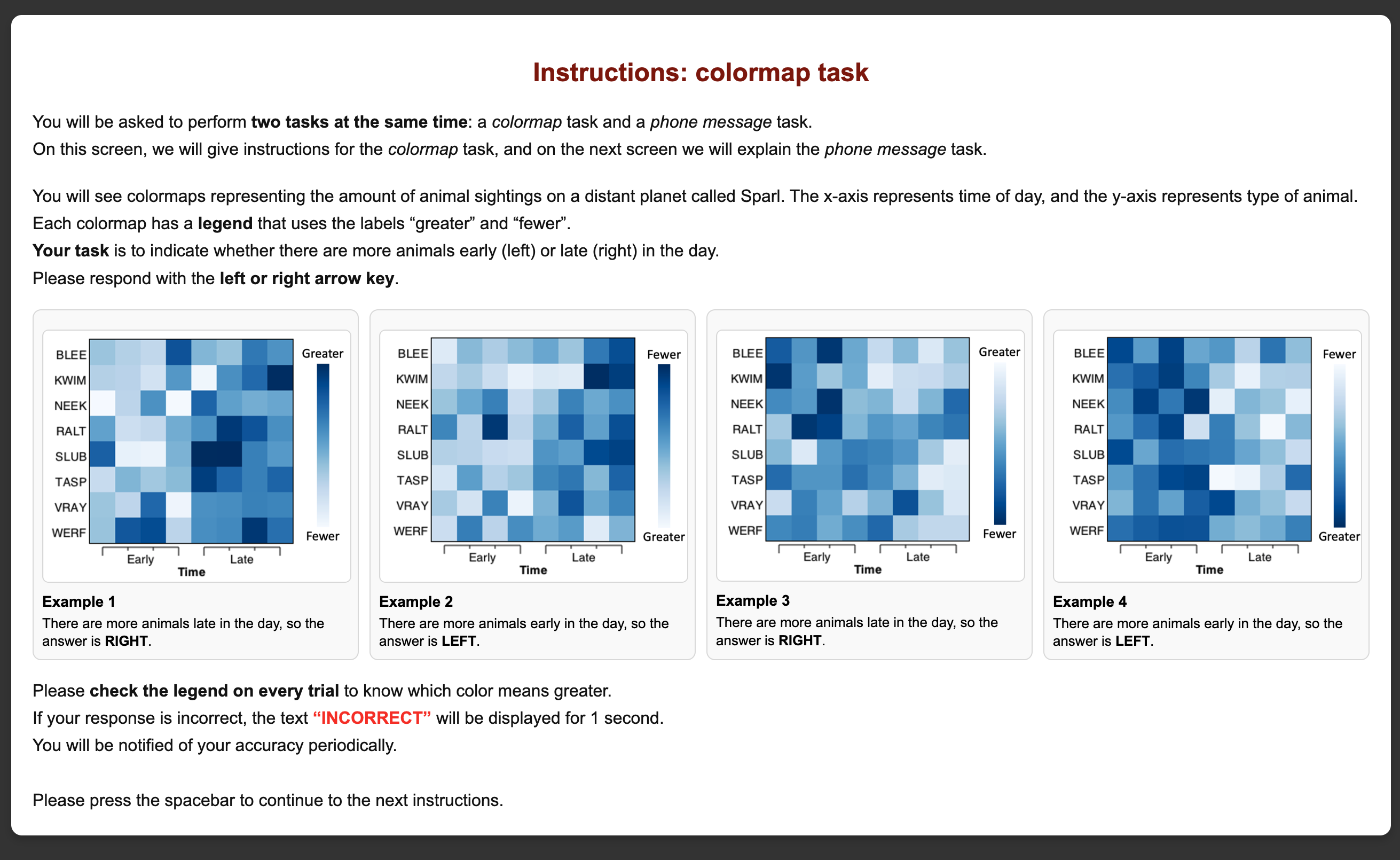}
    \caption{Experiment 1 colormap task instructions for the participants in the dual-task condition.}
    \label{fig:instruction-colormap-dual}
\end{figure}

\begin{figure}[H]
    \centering
    \includegraphics[width=\textwidth]{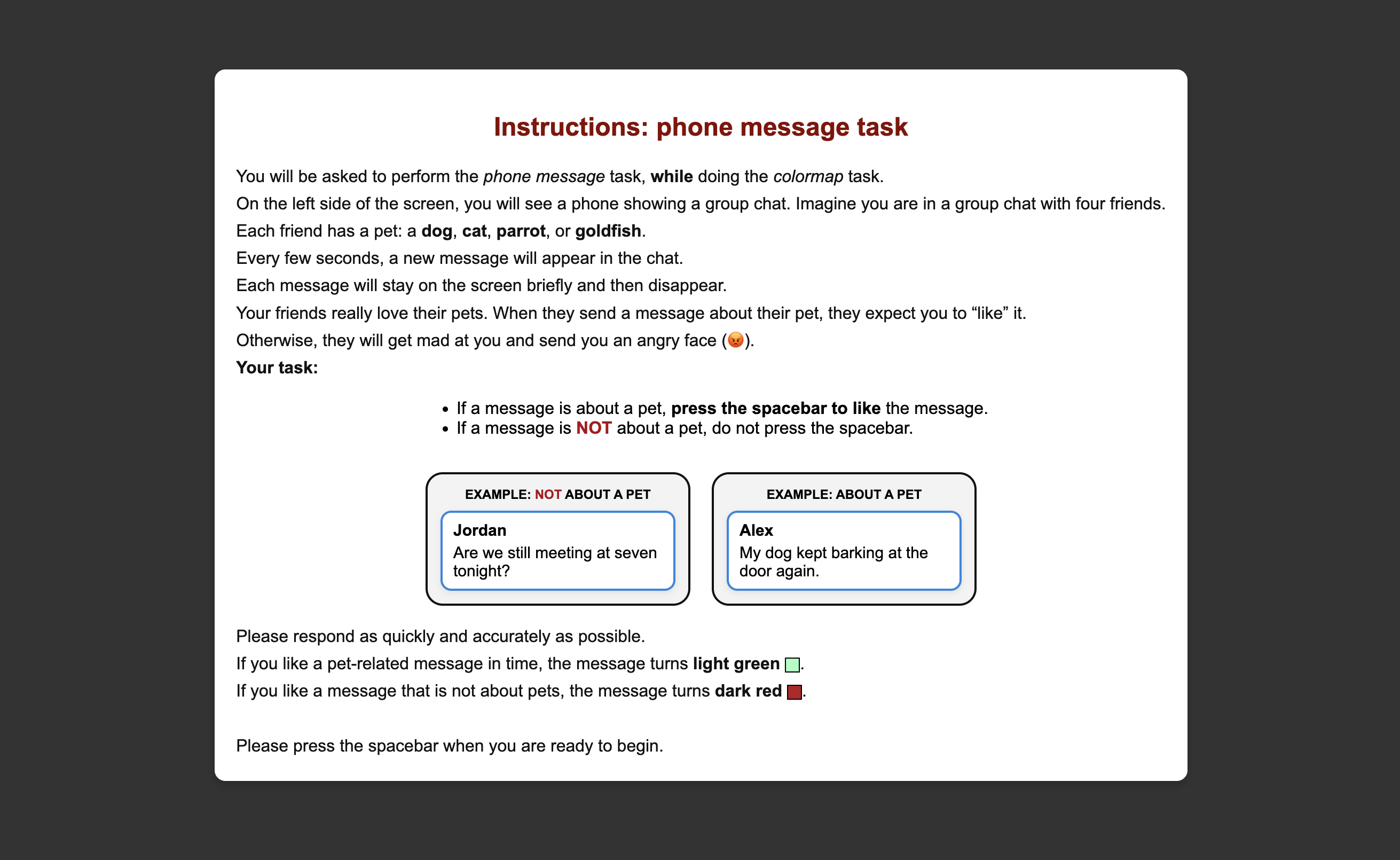}
    \caption{Phone Message Task instructions for the participants in the dual-task condition.}
    \label{fig:instruction-phone}
\end{figure}

\begin{figure}[H]
    \centering
    \includegraphics[width=0.5\textwidth]{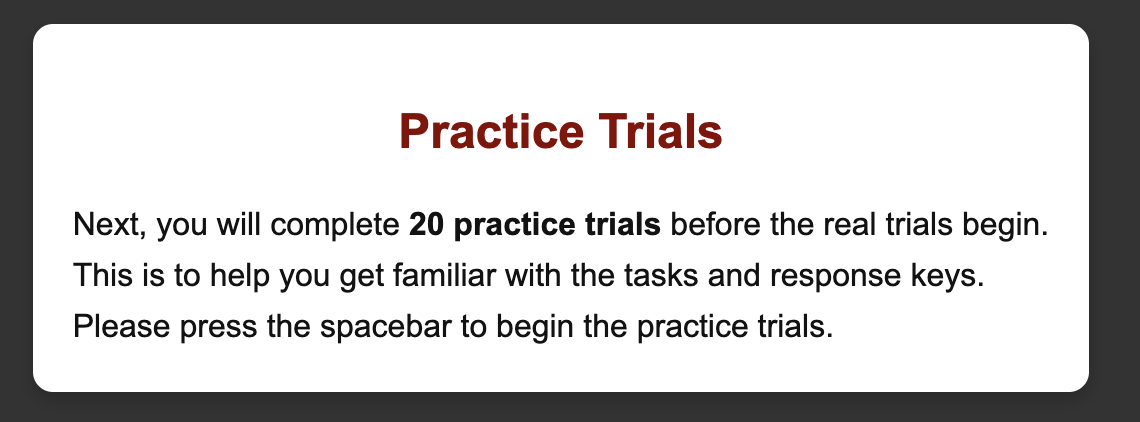}
    \caption{Instruction that pops up right before the start of the practice block.}
    \label{fig:practice}
\end{figure}

\begin{figure}[H]
    \centering
    \includegraphics[width=0.9\textwidth]{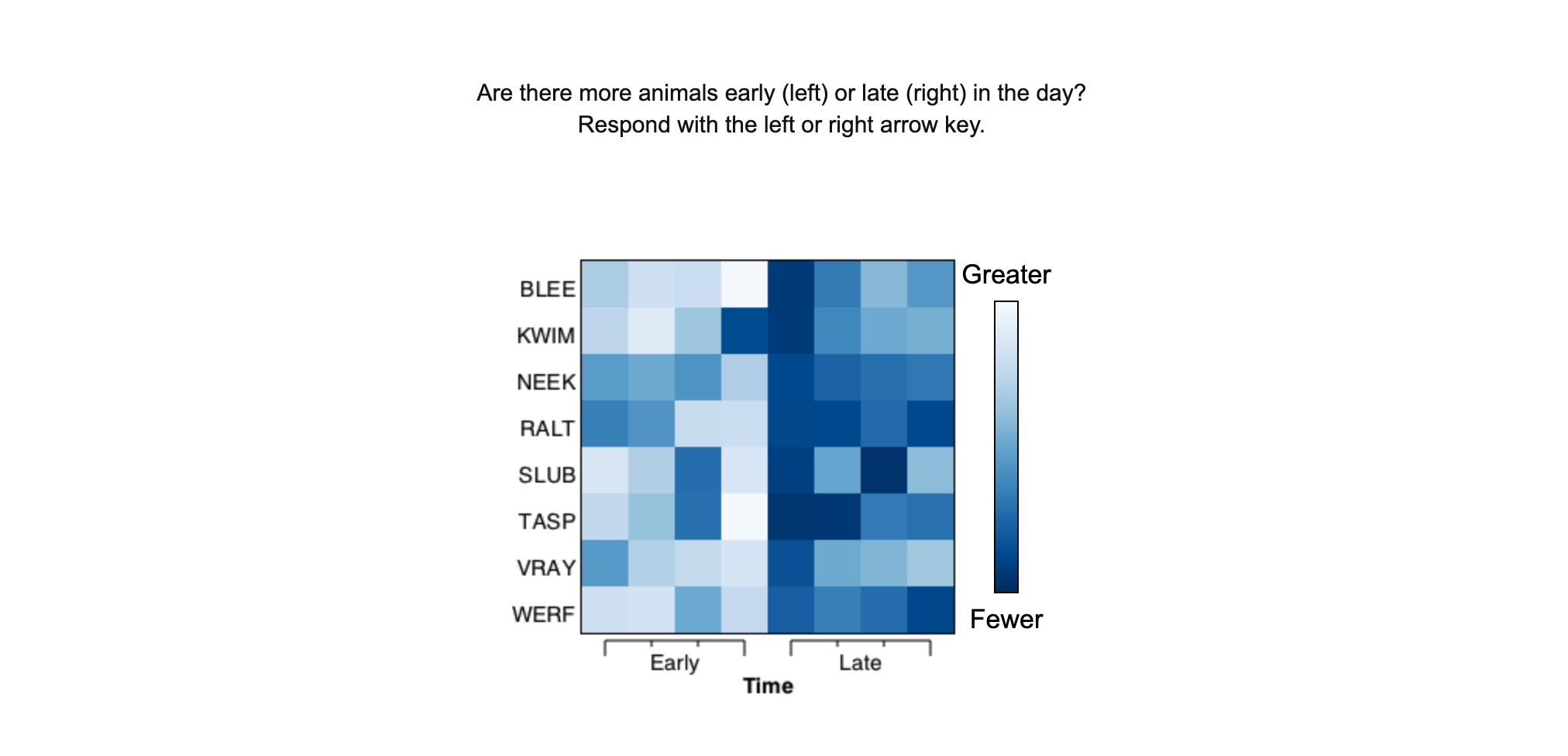}
    \caption{A screenshot of a trial in the single-task condition.}
    \label{fig:trial-single}
\end{figure}

\begin{figure}[H]
    \centering
    \includegraphics[width=0.9\textwidth]{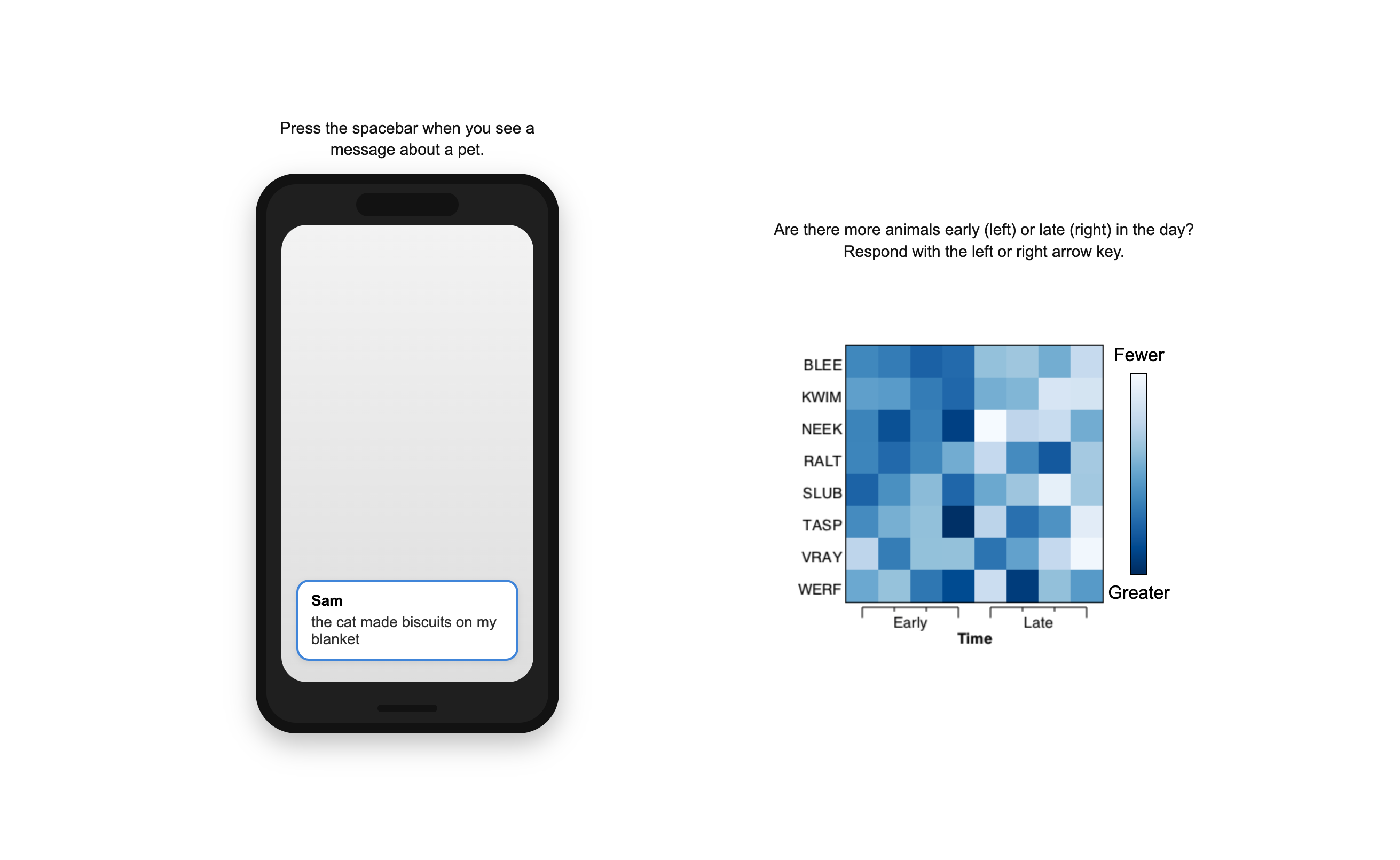}
    \caption{A screenshot of a trial in the dual-task condition.}
    \label{fig:trial-dual}
\end{figure}

The participants in the single-task condition only saw the instruction for the colormap task (Fig. \ref{fig:instruction-colormap-single}). The participants in the dual-task condition saw the instruction for both the colormap task and the phone message task (Fig. \ref{fig:instruction-colormap-dual} and Fig,. \ref{fig:instruction-phone}). Then they completed the practice task, which was prompted by a short message (Fig. \ref{fig:practice}). After practice of 20 trials, participants performed 4 blocks of 20 experiment trials. The trials looked like Fig. \ref{fig:trial-single} or Fig. \ref{fig:trial-dual} depending on the task condition. 

\vspace{-2em}

\vspace{-2em}

\FloatBarrier

\subsection{Experiment 2 Instructions}
\label{sec:instructions_exp2}

\begin{figure}[H]
    \centering
    \includegraphics[width=0.92\linewidth]{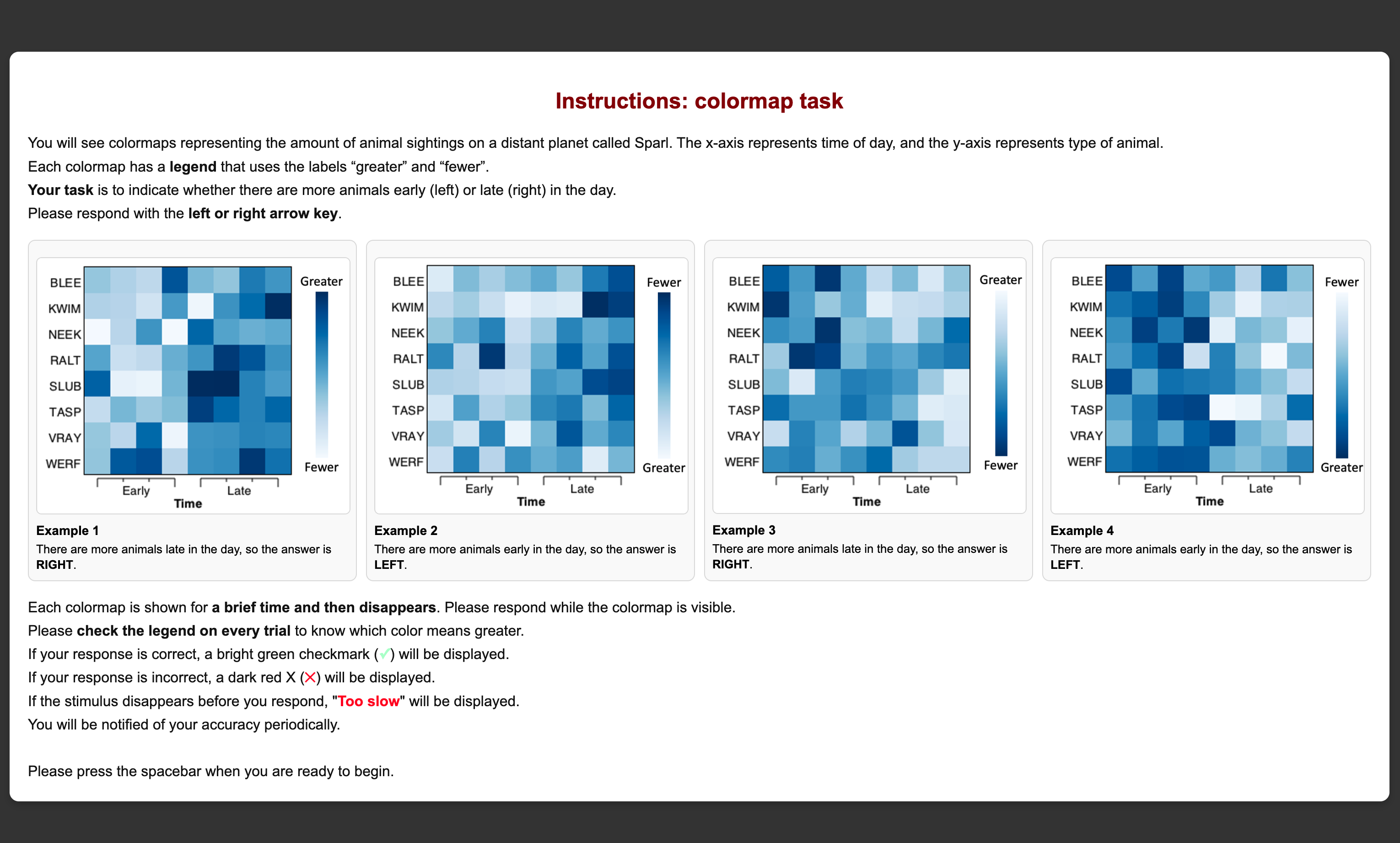}
    \caption{Experiment 2 colormap task instructions for the participants in the single-task condition.}
    \label{fig:exp2-instruction-colormap-single}
\end{figure}

\begin{figure}[H]
    \centering
    \includegraphics[width=0.92\linewidth]{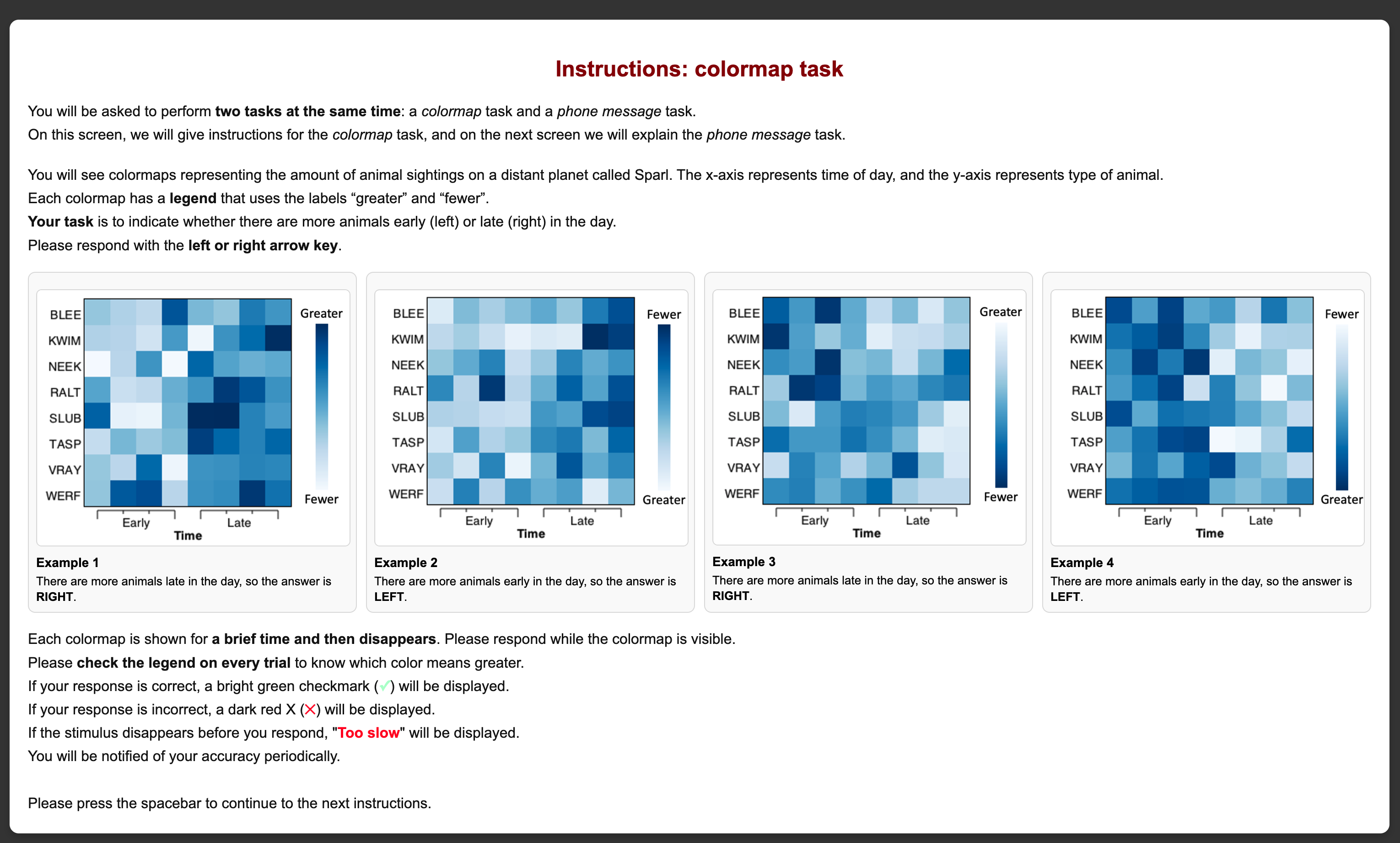}
    \caption{Experiment 2 colormap task instructions for the participants in the dual-task condition.}
    \label{fig:exp2-instruction-colormap-dual}
\end{figure}

The participants in the single-task condition only saw instructions for the colormap task (Fig. \ref{fig:exp2-instruction-colormap-single}) while participants in the dual-task condition also saw instructions for the phone message task (Fig. \ref{fig:exp2-instruction-colormap-dual} and Fig,. \ref{fig:instruction-phone}). The colormap task instructions were slightly different from that of Experiment 1 due to the added time constraint. Experiment 2 was the same as Experiment 1 except each trial moved on automatically if the participant did not answer within the time constraint and they received a "Too slow" message. 

\vspace{-.5
em}

\vspace{-1em}

\FloatBarrier

\subsection{Experiment 1 Analysis}

We analyzed the data using the lmerTest package in R to fit the logistic mixed-effects regression model predicting Accuracy and the linear mixed-effects regression model predicting the Response Time (RT) of accurate trials. The tables in this section report the output from each fitted model.

We used a significance criterion of $p < .05$ to determine whether each fixed effect from the linear mixed-effects regression model (for RT) and the logistic mixed-effects regression model (for Accuracy) was statistically significant. When models did not converge, we simplified the random-effects structure by iteratively removing random slopes until the model converged.

We first present our results discussed in the main body of the paper and then the results for the unpruned data for transparency. As mentioned in `Measures and Exclusions' for Section \ref{sec:exp-1-methods} the pruning procedure was unintentionally omitted from the preregistration, and therefore, the pruned analyses with regard to our preregistered hypotheses are exploratory. However, we present the pruned data analyses in the main paper because participants had unlimited time, and extremely long RTs may reflect disengagement rather than the decision process of interest.


\subsubsection{Accuracy (pruned)}

\begin{table}[H]
\small	
   \centering
   \caption{Experiment 1 full logistic mixed-effects regression model results on accuracy (pruned).}
   \begin{tabular}{llllcccc}
   \hline
        \textbf{Model} &  &  & \textbf{Predictor} & \textbf{$\beta$} & \textbf{SE} & \textbf{$z$} & \textbf{$p$}  \\
            \hline
            Full & & & Intercept  & 3.809  &  0.114 & 33.307 & $<.001$ \\
            & & & Height  & 0.214  &  0.102 & 2.099 & $.036$ \\
            & & & Lightness  & 0.185  &  0.115 & 1.612 & $.107$ \\
            & & & Task  & -0.802  &  0.229 & -3.507 & $<.001$ \\
            & & & Height:Lightness  & 0.280  &  0.193 & 1.451 & $.147$ \\
            & & & Height:Task  & 0.085  &  0.204 & 0.416 & $.678$ \\
            & & & Lightness:Task  & 0.460  &  0.229 & 2.009 & $.045$ \\
            & & & Height:Lightness:Task  & 0.075  &  0.385 & 0.194 & $.846$ \\
            \hline
            & Dual &  & Intercept  & 3.364  &  0.162 & 20.732 & $<.001$ \\
            & & & Height  & 0.361  &  0.108 & 3.333 & $<.001$ \\
            & & & Lightness  & 0.554  &  0.108 & 5.118 & $<.001$ \\
            & & & Height:Lightness  & 0.573  &  0.216 & 2.647 & $.008$ \\
            \hline
             & Single &  & Intercept  & 4.225  &  0.203 & 20.849 & $<.001$ \\
            & & & Height  & 0.293  &  0.175 & 1.674 & $.094$ \\
            & & & Lightness  & 0.126  &  0.171 & 0.738 & $.461$ \\
            & & & Height:Lightness  & 0.476  &  0.302 & 1.577 & $.115$ \\
\hline
   \end{tabular}

   \label{tab:exp1_results_acc}
\end{table}

\textbf{Models}

Base model:
\begin{center}
correct $\sim$ task$*$lightness$*$height + (1 + lightness$*$height $|$ participant)
\end{center}

To unpack the interaction between task and lightness mapping, we ran our model within each task condition.

\vspace{0.5em}
Base model, within dual-task:
\begin{center}
    correct $\sim$ lightness$*$height + (1 $|$ participant)
\end{center}

Base model, within single-task:
\begin{center}
    correct $\sim$ lightness$*$height + (1 + lightness + height $||$ participant)
\end{center}







\FloatBarrier

\subsubsection{Accuracy (unpruned)}

\begin{figure}[H]
    \centering
    \includegraphics[width=0.7\linewidth]{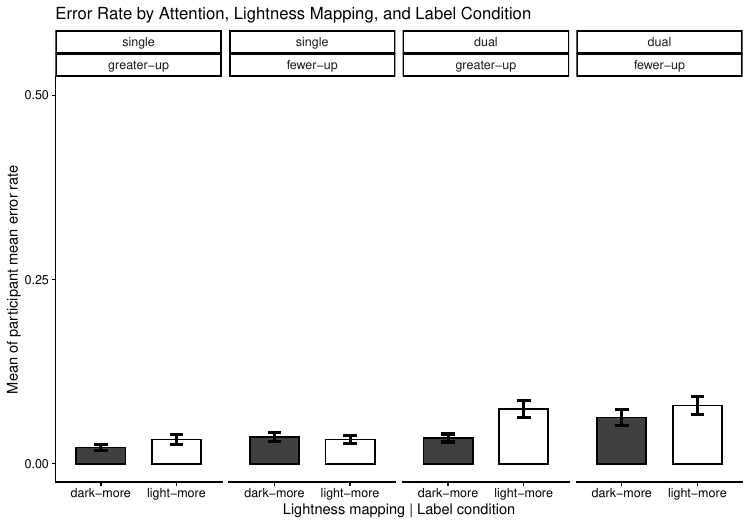}
    \caption{Error rate in the unpruned dataset. Bars indicate mean of participant accuracy.}
    \label{fig:exp1_acc_unpruned}
\end{figure}

\begin{table}[H]
\small	

   \centering
   \caption{Experiment 1 full logistic mixed-effects regression model results on accuracy (unpruned).}
   \begin{tabular}{llllcccc}
   \hline
        \textbf{Model} &  &  & \textbf{Predictor} & \textbf{$\beta$} & \textbf{SE} & \textbf{$z$} & \textbf{$p$}  \\
            \hline
            Full &  &  & Intercept & 3.808 & 0.116 & 32.949 & $<.001$ \\
             &  &  & Height & 0.250 & 0.101 & 2.482 & $.013$ \\
             &  &  & Lightness & 0.181 & 0.114 & 1.595 & $.111$ \\
             &  &  & Task & -0.783 & 0.231 & -3.387 & $<.001$ \\
             &  &  & Height:Lightness & 0.326 & 0.190 & 1.716 & $.086$ \\
             &  &  & Height:Task & 0.072 & 0.202 & 0.357 & $.721$ \\
             &  &  & Lightness:Task & 0.478 & 0.228 & 2.101 & $.036$ \\
             &  &  & Height:Lightness:Task & 0.050 & 0.380 & 0.132 & $.895$ \\
            \hline
             & Dual &  & Intercept & 3.368 & 0.163 & 20.639 & $<.001$ \\
             &  &  & Height & 0.367 & 0.108 & 3.397 & $<.001$ \\
             &  &  & Lightness & 0.565 & 0.108 & 5.229 & $<.001$ \\
             &  &  & Height:Lightness & 0.582 & 0.216 & 2.696 & $.007$ \\
            \hline
             & Single &  & Intercept & 4.221 & 0.205 & 20.582 & $<.001$ \\
             &  &  & Height & 0.311 & 0.173 & 1.803 & .071 \\
             &  &  & Lightness & 0.122 & 0.173 & 0.709 & .478 \\
             &  &  & Height:Lightness & 0.514 & 0.298 & 1.728 & .084 \\

\hline
   \end{tabular}

   \label{tab:exp1_results_acc_unpruned}
\end{table}

\vspace{0.5em}
\textbf{Models}

Base model:
\begin{center}
correct $\sim$ task$*$lightness$*$height + (1 + lightness$*$height $|$ participant)
\end{center}

To unpack the interaction between task and lightness mapping, we ran our model within each task condition.

\vspace{0.5em}
Base model, within single-task:
\begin{center}
    correct $\sim$ lightness$*$height + (1 + lightness + height $||$ participant)
\end{center}

Base model, within dual-task:
\begin{center}
    correct $\sim$ lightness$*$height + (1 $|$ participant)
\end{center}




\FloatBarrier

\subsubsection{Response time (pruned)}
There were 109 trials out of 13360 excluded from pruning the outlier trials of each participant. See \ref{sec:exp1_pruned_by_participant} of this document to see which trials were pruned out per participant.

\begin{table}[H]
\small	

   \centering
   \caption{Experiment 1 full linear mixed-effects regression model results on pruned RT data.}
   \begin{tabular}{llllccccc}
   \hline
        \textbf{Model} &  &  & \textbf{Predictor} & \textbf{$\beta$} & \textbf{df} & \textbf{SE} & \textbf{$t$} & \textbf{$p$}  \\
            \hline
            Full &  &  & Intercept & 1999.957 & 164.907 & 58.983 & 33.907 & $<.001$ \\
             &  &  & Height & -82.183 & 12308.517 & 18.987 & -4.328 & $<.001$ \\
             &  &  & Lightness & -170.878 & 159.283 & 22.655 & -7.542 & $<.001$ \\
             &  &  & Task & 377.518 & 164.907 & 117.967 & 3.200 & $.002$ \\
             &  &  & Height:Lightness & -205.738 & 12310.840 & 37.971 & -5.418 & $<.001$ \\
             &  &  & Height:Task & -82.051 & 12308.517 & 37.973 & -2.161 & $.031$ \\
             &  &  & Lightness:Task & -52.867 & 159.283 & 45.311 & -1.167 & $.245$ \\
             &  &  & Height:Lightness:Task & -52.876 & 12310.840 & 75.941 & -0.696 & $.486$ \\
            \hline
             & High-more &  & Intercept & 1958.116 & 164.755 & 59.987 & 32.642 & $<.001$ \\
             &  &  & Lightness & -273.140 & 6195.205 & 27.067 & -10.091 & $<.001$ \\
             &  &  & Task & 336.487 & 164.755 & 119.974 & 2.805 & $.006$ \\
             &  &  & Lightness:Task & -81.133 & 6195.205 & 54.133 & -1.499 & $.134$ \\
            \hline
            & Low-more &  & Intercept & 2041.334 & 164.903 & 59.112 & 34.533 & $<.001$ \\
             &  &  & Lightness & -66.508 & 6109.894 & 26.881 & -2.474 & $.013$ \\
             &  &  & Task & 417.583 & 164.903 & 118.224 & 3.532 & $<.001$ \\
             &  &  & Lightness:Task & -22.228 & 6109.894 & 53.762 & -0.413 & $.679$ \\
            \hline
             & Dual &  & Intercept & 2188.382 & 81.931 & 77.611 & 28.197 & $<.001$ \\
             &  &  & Height & -123.144 & 6037.737 & 24.380 & -5.051 & $<.001$ \\
             &  &  & Lightness & -197.056 & 79.045 & 26.256 & -7.505 & $<.001$ \\
             &  &  & Height:Lightness & -231.875 & 6038.883 & 48.756 & -4.756 & $<.001$ \\
            \hline
             & Single &  & Intercept & 1811.254 & 82.954 & 88.682 & 20.424 & $<.001$ \\
             &  &  & Height & -41.141 & 6271.303 & 28.931 & -1.422 & $.155$ \\
             &  &  & Lightness & -144.595 & 79.700 & 36.654 & -3.945 & $<.001$ \\
             &  &  & Height:Lightness & -179.394 & 6272.408 & 57.859 & -3.101 & $.002$ \\

\hline
   \end{tabular}

   \label{tab:exp1_results_rt_pruned}
\end{table}

\vspace{0.5em}
\textbf{Models}

Base model:
\begin{center}
correct $\sim$ task$*$lightness$*$height + (1 + lightness $||$ participant)
\end{center}

To unpack the task and height mapping interaction by we ran our model within each task condition.

\vspace{0.5em}
Base model, within dual-task:
\begin{center}
RT $\sim$ lightness$*$height + (1 + lightness $||$ participant)
\end{center}

Base model, within single-task:
\begin{center}
RT $\sim$ lightness$*$height + (1 + lightness $||$ participant)
\end{center}







To unpack the interaction between height mapping and lightness mapping we ran our model within each height mapping.

\vspace{0.5em}
Base model, within high-more mapping:
\begin{center}
RT $\sim$ task$*$lightness + (1 $|$ participant)
\end{center}

Base model, within low-more mapping:
\begin{center}
RT $\sim$ task$*$lightness + (1 $|$ participant)
\end{center}

\FloatBarrier

\subsubsection{Response time (unpruned)}
\label{sec:exp1_rt_unpruned_analysis}

\begin{figure}[H]
    \centering
    \includegraphics[width=0.7\linewidth]{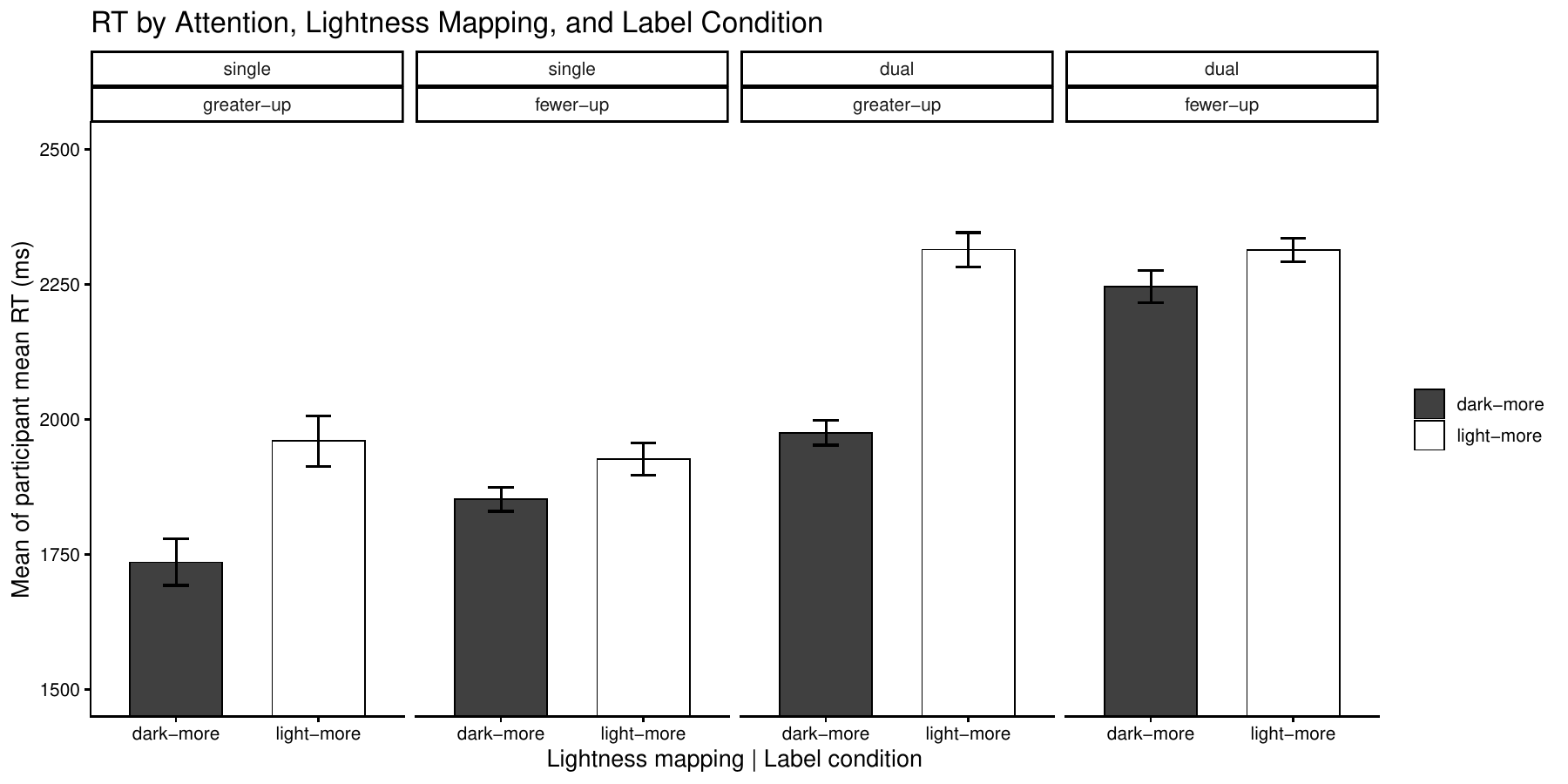}
    \caption{RT of correct trials in the unpruned dataset. Bars indicate mean of participant means. Error bars account for within-participant variance (Cousineau-Morey method~\cite{morey2008confidence}).}
    \label{fig:exp1_rt_unpruned}
\end{figure}

\begin{table}[H]
\small	

   \centering
   \caption{Experiment 1 full linear mixed-effects regression model results on unpruned RT data.}
   \begin{tabular}{llllccccc}
   \hline
        \textbf{Model} &  &  & \textbf{Predictor} & \textbf{$\beta$} & \textbf{df} & \textbf{SE} & \textbf{$t$} & \textbf{$p$}  \\
            \hline
            Full &  &  & Intercept & 2038.582 & 164.851 & 59.463 & 34.283 & $<.001$ \\
             &  &  & Height & -88.145 & 12402.015 & 22.114 & -3.986 & $<.001$ \\
             &  &  & Lightness & -174.292 & 158.672 & 26.480 & -6.582 & $<.001$ \\
             &  &  & Task & 342.976 & 164.851 & 118.926 & 2.884 & $.004$ \\
             &  &  & Height:Lightness & -206.494 & 12404.878 & 44.226 & -4.669 & $<.001$ \\
             &  &  & Height:Task & -91.664 & 12402.015 & 44.229 & -2.073 & $.038$ \\
             &  &  & Lightness:Task & -54.222 & 158.672 & 52.960 & -1.024 & $.307$ \\
             &  &  & Height:Lightness:Task & -117.982 & 12404.878 & 88.451 & -1.334 & $.182$ \\
            \hline
             & High-more &  & Intercept & 1994.579 & 164.637 & 60.440 & 33.001 & $<.001$ \\
             &  &  & Lightness & -280.651 & 163.188 & 44.113 & -6.362 & $<.001$ \\
             &  &  & Task & 296.300 & 164.267 & 120.881 & 2.451 & $.015$ \\
             &  &  & Lightness:Task & -114.958 & 163.188 & 88.226 & -1.303 & $.194$ \\
            \hline
            & Low-more &  & Intercept & 2082.715 & 164.768 & 60.425 & 34.468 & $<.001$ \\
             &  &  & Lightness & -69.462 & 6162.178 & 30.848 & -2.252 & $.024$ \\
             &  &  & Task & 388.620 & 164.768 & 120.849 & 3.216 & $.002$ \\
             &  &  & Lightness:Task & 9.417 & 6162.178 & 61.696 & 0.153 & $.879$ \\
            \hline
             & Dual &  & Intercept & 2209.962 & 81.917 & 78.982 & 27.981 & $<.001$ \\
             &  &  & Height & -134.025 & 76.171 & 28.424 & -4.715 & $<.001$ \\
             &  &  & Lightness & -201.272 & 78.466 & 30.757 & -6.544 & $<.001$ \\
             &  &  & Height:Lightness & -265.517 & 5999.709 & 54.382 & -4.882 & $<.001$ \\
            \hline
             & Single &  & Intercept & 1866.747 & 82.910 & 88.697 & 21.046 & $<.001$ \\
             &  &  & Height & -42.384 & 6425.166 & 34.655 & -1.223 & $.221$ \\
             &  &  & Lightness & -146.288 & 6425.221 & 34.657 & -4.221 & $<.001$ \\
             &  &  & Height:Lightness & -147.189 & 6425.064 & 69.304 & -2.124 & $.034$ \\

\hline
   \end{tabular}

   \label{tab:exp1_results_rt_unpruned}
\end{table}

\vspace{0.5em}
\textbf{Models}

Base model:
\begin{center}
RT $\sim$ task$*$lightness$*$height + (1 + lightness $||$ participant)
\end{center}

To unpack the interaction between task and height mapping, we ran our model within each task condition. 

\vspace{0.5em}
Base model, within dual-task:
\begin{center}
RT $\sim$ lightness$*$height + (1 + lightness + height $||$ participant)
\end{center}

Base model, within single-task:
\begin{center}
RT $\sim$ lightness$*$height + (1 $|$ participant)
\end{center}








To unpack the interaction between lightness mapping and height mapping, we ran our model within each height mapping condition.

\vspace{0.5em}
Base model, within high-more mapping:
\begin{center}
RT $\sim$ task$*$lightness + (1 + lightness $|$ participant)
\end{center}

Base model, within low-more mapping:
\begin{center}
RT $\sim$ task$*$lightness + (1 $|$ participant)
\end{center}

\FloatBarrier

\subsection{Experiment 2 Analysis}

\subsubsection{Miss-rate}

\begin{table}[H]
\small	

   \centering
   \caption{Experiment 2 full logistic mixed-effects regression model results for miss rate.}
   \begin{tabular}{llllcccc}
   \hline
        \textbf{Model} &  &  & \textbf{Predictor} & \textbf{$\beta$} & \textbf{SE} & \textbf{$z$} & \textbf{$p$}  \\
            \hline
            Full &  &  & Intercept & -3.527 & 0.130 & -27.140 & $.001$ \\
             &  &  & Height & -0.158 & 0.160 & -0.987 & $.324$ \\
             &  &  & Lightness & -0.405 & 0.158 & -2.560 & $.010$ \\
             &  &  & Task & 1.829 & 0.245 & 7.480 & $<.001$ \\
             &  &  & Height:Lightness & 0.085 & 0.242 & 0.353 & $.724$ \\
             &  &  & Height:Task & -0.179 & 0.259 & -0.691 & $.489$ \\
             &  &  & Lightness:Task & 0.300 & 0.251 & 1.194 & $.233$ \\
             &  &  & Height:Lightness:Task & -0.963 & 0.467 & -2.063 & $.039$ \\
            \hline
             & Dual &  & Intercept & -2.587 & 0.137 & -18.932 & $<.001$ \\
             &  &  & Height & -0.191 & 0.118 & -1.622 & $.105$ \\
             &  &  & Lightness & -0.191 & 0.118 & -1.616 & $.106$ \\
             &  &  & Height:Lightness & -0.436 & 0.206 & -2.115 & $.034$ \\
             \hline
             &  & High-more & Intercept & -2.739 & 0.169 & -16.226 & $<.001$ \\
             &  &  & Lightness & -0.423 & 0.168 & -2.526 & $.012$ \\
            \hline
             &  & Low-more & Intercept & -2.448 & 0.143 & -17.102 & $<.001$ \\
             &  &  & Lightness & 0.041 & 0.140 & 0.292 & $.771$ \\
             \hline
             &  & Dark-more & Intercept & -2.751 & 0.170 & -16.162 & $<.001$ \\
             &  &  & Height & -0.423  &  0.166 & -2.556 & $.011$ \\
            \hline
             &  & Light-more & Intercept  & -2.484  &  0.151 & -16.475 & $<.001$ \\
             &  &  & Height & 0.043  &  0.147 & 0.294 & $.769$ \\
            \hline
             & Single &  & Intercept & -4.562 & 0.253 & -18.029 & $<.001$ \\
             &  &  & Height & 0.047 & 0.208 & 0.225 & $.822$ \\
             &  &  & Lightness & -0.423 & 0.208 & -2.030 & $.042$ \\
             &  &  & Height:Lightness & 0.515 & 0.416 & 1.237 & $.216$ \\

\hline
   \end{tabular}

   \label{tab:exp2_results_miss}
\end{table}

\vspace{0.5em}
\textbf{Models}

Base model:
\begin{center}
missed $\sim$ task$*$lightness$*$height + (1 + lightness + height $|$ participant)
\end{center}

To unpack the 3-way interaction we ran our model within each task condition, and then within each height mapping.

\vspace{0.5em}
Base model, within dual-task:
\begin{center}
missed $\sim$ lightness$*$height + (1 + lightness + height $||$ participant)
\end{center}

Base model, within single-task:
\begin{center}
missed $\sim$ lightness$*$height + (1 $|$ participant)
\end{center}




\FloatBarrier

\subsubsection{Response time}

\begin{table}[H]
\small	

   \centering
   \caption{Experiment 1 full linear mixed-effects regression model results.}
   \begin{tabular}{llllccccc}
   \hline
        \textbf{Model} &  &  & \textbf{Predictor} & \textbf{$\beta$} & \textbf{df} & \textbf{SE} & \textbf{$t$} & \textbf{$p$}  \\
            \hline
            Full &  &  & Intercept & 1026.160 & 113.937 & 20.521 & 50.006 & $<.001$ \\
             &  &  & Height & -40.725 & 103.334 & 7.902 & -5.154 & $<.001$ \\
             &  &  & Lightness & -48.054 & 80.984 & 6.417 & -7.489 & $<.001$ \\
             &  &  & Task & 86.493 & 113.937 & 41.041 & 2.107 & $.037$ \\
             &  &  & Height:Lightness & -21.250 & 6364.476 & 11.648 & -1.824 & $.068$ \\
             &  &  & Height:Task & 27.432 & 103.334 & 15.803 & 1.736 & $.086$ \\
             &  &  & Lightness:Task & -1.464 & 80.984 & 12.834 & -0.114 & $.909$ \\
             &  &  & Height:Lightness:Task & -30.957 & 6364.476 & 23.295 & -1.329 & $.184$ \\

\hline
   \end{tabular}

   \label{tab:exp2_results_rt}
\end{table}

\vspace{0.5em}
\textbf{Models}

Base model:
\begin{center}
RT $\sim$ task$*$lightness$*$height + (1 + lightness + height $||$ participant)
\end{center}

\FloatBarrier

\subsubsection{Accuracy}

\begin{table}[H]
\small	

   \centering
   \caption{Experiment 2 full logistic mixed-effects regression model results.}
   \begin{tabular}{llllcccc}
   \hline
        \textbf{Model} &  &  & \textbf{Predictor} & \textbf{$\beta$} & \textbf{SE} & \textbf{$z$} & \textbf{$p$}  \\
            \hline
            Full &  &  & Intercept & 1.270 & 0.127 & 9.974 & $<.001$ \\
             &  &  & Height & 0.315 & 0.074 & 4.243 & $<.001$ \\
             &  &  & Lightness & 0.855 & 0.137 & 6.230 & $<.001$ \\
             &  &  & Task & -1.151 & 0.254 & -4.529 & $<.001$ \\
             &  &  & Height:Lightness & 0.551 & 0.111 & 4.971 & $<.001$ \\
             &  &  & Height:Task & -0.027 & 0.148 & -0.184 & $.854$ \\
             &  &  & Lightness:Task & 0.269 & 0.274 & 0.982 & $.326$ \\
             &  &  & Height:Lightness:Task & 0.108 & 0.222 & 0.486 & $.627$ \\
            \hline
             & High-more &  & Intercept & 1.443 & 0.134 & 10.743 & $<.001$ \\
             &  &  & Lightness & 1.167 & 0.147 & 7.955 & $<.001$ \\
             &  &  & Task & -1.111 & 0.267 & -4.167 & $<.001$ \\
             &  &  & Lightness:Task & 0.333 & 0.292 & 1.143 & $.253$ \\
            \hline
            & Low-more &  & Intercept & 1.131 & 0.135 & 8.373 & $<.001$ \\
             &  &  & Lightness & 0.572 & 0.156 & 3.680 & $<.001$ \\
             &  &  & Task & -1.151 & 0.269 & -4.275 & $<.001$ \\
             &  &  & Lightness:Task & 0.202 & 0.311 & 0.651 & $.515$ \\

\hline
   \end{tabular}

   \label{tab:exp2_results_acc}
\end{table}

\vspace{0.5em}
\textbf{Models}

Base model:
\begin{center}
correct $\sim$ task$*$lightness$*$height + (1 + lightness + height $||$ participant)
\end{center}

To unpack the interaction between height and lightness mapping we ran our model within each height condition.

\vspace{0.5em}
Base model, within high-more mapping:
\begin{center}
    correct $\sim$ task$*$lightness + (1 + lightness $||$ participant)
\end{center}

Base model, within low-more mapping:
\begin{center}
    correct $\sim$ task$*$lightness + (1 + lightness $||$ participant)
\end{center}

\FloatBarrier

\subsection{LBA}

We followed the modeling approach outlined by Donkin et al. (2011). We first specified a maximally complex free model consistent with our a priori assumptions. We then examined the fitted parameter values to identify opportunities for simplifying the model without compromising its ability to capture the data. Based on these observations, we constructed and fit a set of candidate models with reduced complexity, and selected the final model according to predefined evaluation criteria.

\subsubsection{Experiment 1 LBA models}

For Experiment 1 we used the pruned dataset to fit the LBA models.

\vspace{0.5em}
\noindent\textbf{Free model after a priori assumptions}
\label{sec:exp1_free}

\begin{figure}[H]
    \centering
    \includegraphics[width=0.7\linewidth]{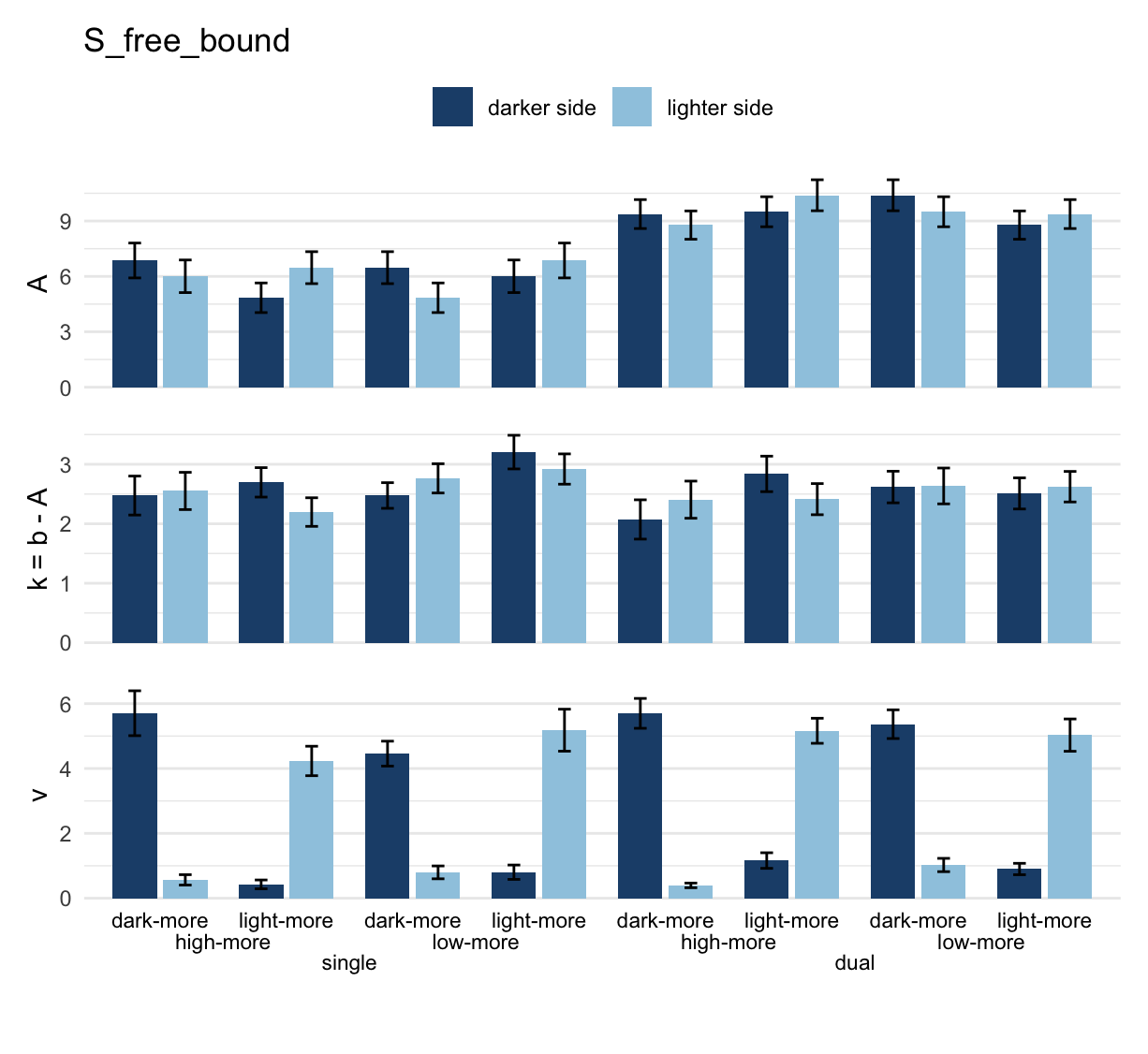}
    \caption{Parameters after fitting the free model to each participant's data. Bars show the mean of parameter values across participants. Error bars show standard error.}
    \label{fig:lba_exp1_free}
\end{figure}

Figure~\ref{fig:lba_exp2_free} shows the parameter values of maximally complex LBA model following the a priori assumptions, fit to each participant. In this model, $A$ varies by \textit{accumulator color} and \textit{legend-top color}. $b$ varies by \textit{lightness mapping}, \textit{height mapping}, and \textit{accumulator color}. $v$ varies by \textit{lightness mapping}, \textit{height mapping}, and \textit{accumulator correctness}.
This means that the total number of parameters fit per participant is $2^2 + 2^3 + 2^3 = 20$. Note that each participant is only within one task condition.

This analysis allowed us to additionally 1) tie $A$ across all conditions except across task 2) tie $k$ across accumulator color (i.e., have same $k$ for the darker side and lighter side accumulator in a given trial).

\FloatBarrier

\vspace{0.5em}
\noindent\textbf{Candidate model evaluation}

There were a total of 8 candidate models. Table~\ref{tab:lba_exp2_candidates} shows the model specs of the 8 candidate models and their evaluations. Evaluations are based on AIC, BIC, and most importantly, whether simulated data can reproduce the behavioral pattern found in the original data. The target pattern we aimed to reproduce was the interaction between task and height mapping on RT, and the interaction between task and lightness mapping on accuracy. These were the main findings of the Experiment 1 analyses.

We planned that each participant model would produce the same number of trials per condition as the real experiment (i.e., 20 trials for each of 2 lightness mapping $\times$ 2 height mapping = 4 conditions). LMER model was fit to predict RT of simulated data. The model had the same formula used in the actual Experiment 1 analysis. 

However, for Experiment~1, we used more simulated trials than were present in the original experiment to obtain a stable estimate of each model's predicted behavior. Accuracy in Experiment~1 was near ceiling, so the targeted interaction effects corresponded to small differences in error rates and were difficult to reproduce from simulated datasets with only 20 trials per condition per participant. We therefore increased the number of simulated trials in increments of 10 and used 50 trials per condition per participant for model evaluation. This increased simulation size was used only to reduce noise in the model predictions.

The $\beta$ and $p$ in the table refers to the coefficient and the $p$-value associated with the interaction between task and height mapping on RT.

\FloatBarrier

\begin{table}[!h]
\small
    \centering
    \caption{Candidate model specs and their evaluations based on our criteria. The $\beta$ and $p$ are based on LMER analysis on simulated data for RT, and whether there is an interaction between task and height mapping.}
    \begin{tabular}{ccclllll}
    \hline
    \multicolumn{3}{c}{Parameter control}          & \multicolumn{5}{c}{Evaluation}         \\
    \cmidrule(lr){1-3} \cmidrule(lr){4-8}
A: accumulator color & v: lightness mapping      & v: height mapping         & AIC & BIC & $\beta$ & $p$ & Recovery \\
True       & True  & True  & 23392      & 26697     &  0.110  & 0.019         & yes      \\
True       & True  & False & 23134      & 25337     & -0.014  & 0.663         & no       \\
True       & False & True  & 23182      & 25385     &  0.014  & 0.826         & no       \\
False      & True  & True  & 23228      & 26257     &  0.038  & 0.439         & no       \\
True       & False & False & 31234      & 32887     & -0.100  & 0.072         & no       \\
False      & True  & False & 23024      & 24952     &  0.005  & 0.868         & no       \\
False      & False & True  & 23099      & 25027     &  0.117  & 0.023         & yes       \\
False      & False & False & 23074      & 24451     & -0.030  & 0.349         & no      
\end{tabular}
\label{tab:lba_exp1_candidates_rt}
\end{table}

\begin{table}[!h]
\small
    \centering
    \caption{Candidate model specs and their evaluations based on our criteria. The $\beta$ and $p$ are based on GLMER analysis on simulated data for accuracy, and whether there is an interaction between task and lightness mapping.}
    \begin{tabular}{ccclllll}
    \hline
    \multicolumn{3}{c}{Parameter control}          & \multicolumn{5}{c}{Evaluation}         \\
    \cmidrule(lr){1-3} \cmidrule(lr){4-8}
A: accumulator color & v: lightness mapping      & v: height mapping         & AIC & BIC & $\beta$ & $p$ & Recovery \\
True       & True  & True  & 23392      & 26697     & -0.470  & 0.028         & yes      \\
True       & True  & False & 23134      & 25337     & -0.464  & 0.038         & yes       \\
True       & False & True  & 23182      & 25385     & -0.397  & 0.125         & no       \\
False      & True  & True  & 23228      & 26257     & -0.384  & 0.009         & yes      \\
True       & False & False & 31234      & 32887     & -0.077  & 0.377         & no       \\
False      & True  & False & 23024      & 24952     & -0.457  & 0.043         & yes       \\
False      & False & True  & 23099      & 25027     &  0.087  & 0.385         & no       \\
False      & False & False & 23074      & 24451     & 0.152   & 0.126         & no         
\end{tabular}
\label{tab:lba_exp1_candidates_acc}
\end{table}

\FloatBarrier

\begin{figure}[H]
    \centering
    \includegraphics[width=0.7\linewidth]{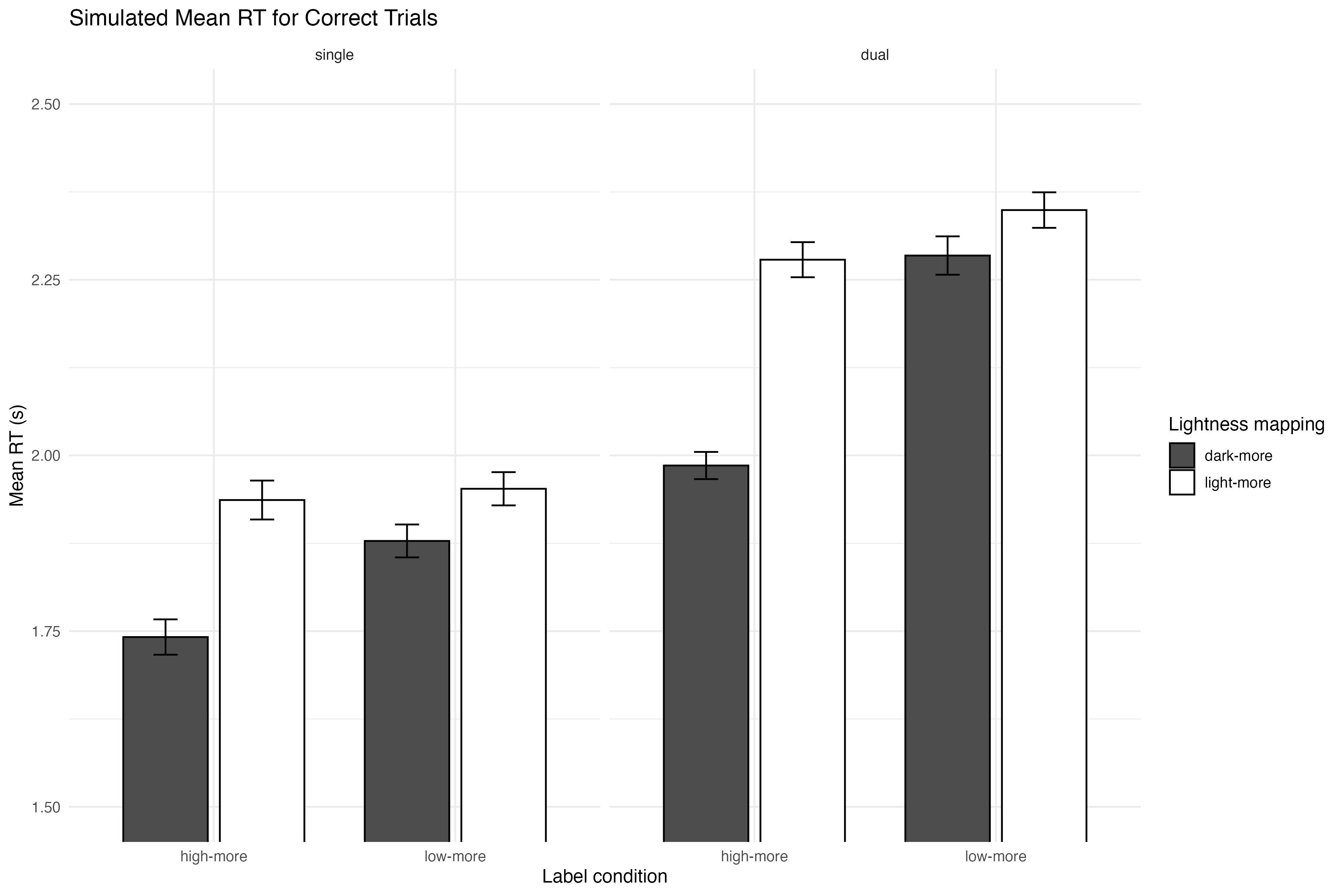}
    \caption{RT from the data simulated by the final model.}
    \label{fig:lba_exp1_simulated_rt}
\end{figure}

\begin{figure}[H]
    \centering
    \includegraphics[width=0.7\linewidth]{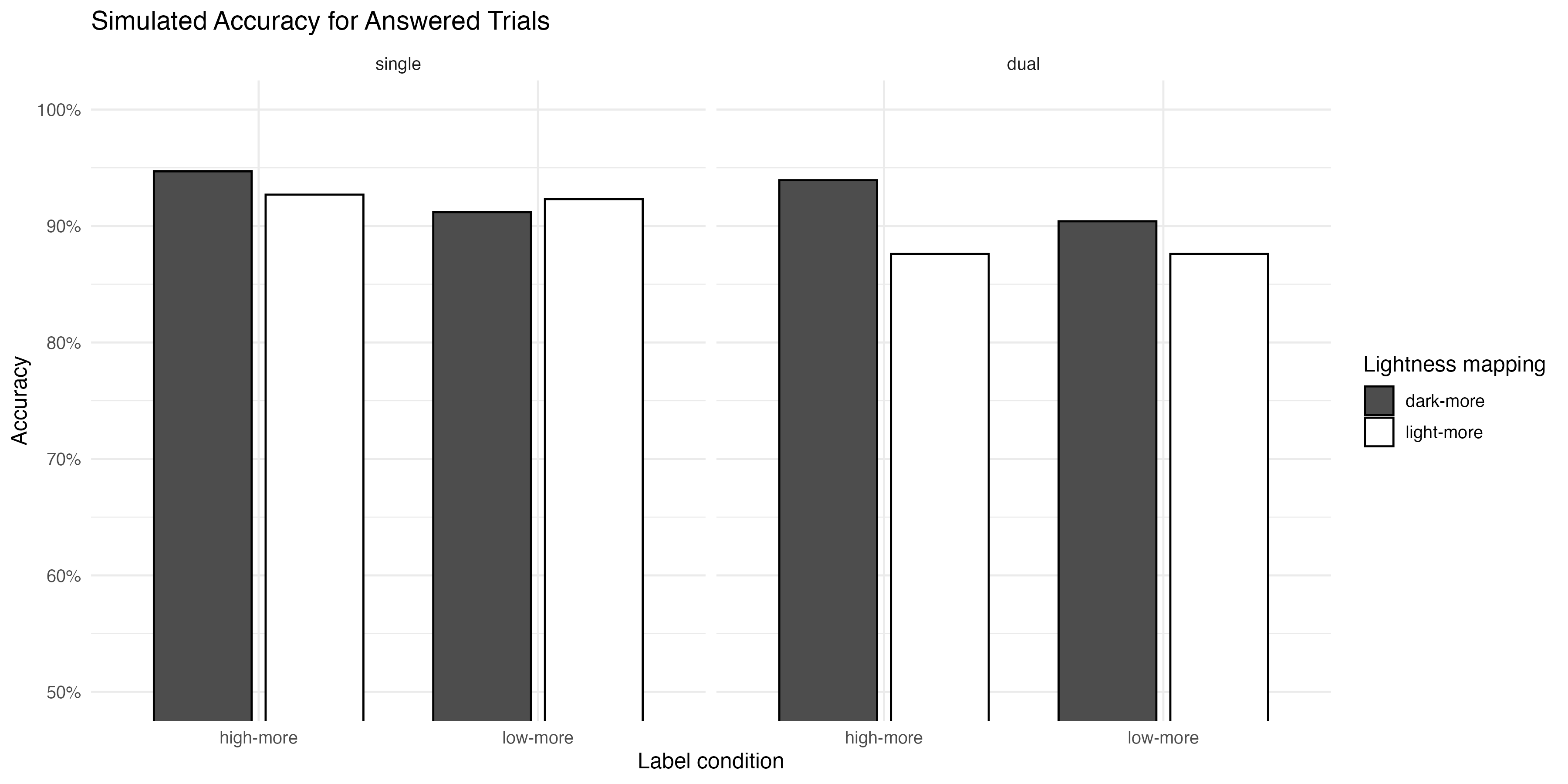}
    \caption{Accuracy from the data simulated by the final model.}
    \label{fig:lba_exp1_simulated_acc}
\end{figure}

The two interactions were both significant in the data simulated by the model corresponding to the first row of Table~\ref{tab:lba_exp1_candidates_acc} and Table~\ref{tab:lba_exp1_candidates_rt}. The pattern of these interactions matched the original data, as seen in Fig. \ref{fig:lba_exp1_simulated_rt} and Fig. \ref{fig:lba_exp1_simulated_acc}. Therefore, we select this model as our final LBA model. The final model is where $A$ varies by accumulator color and $v$ varies by lightness mapping, height mapping, and accumulator correctness.

\FloatBarrier

\vspace{0.5em}
\noindent\textbf{Final model and Parameter Analysis}

\begin{figure}[H]
    \centering
    \includegraphics[width=0.7\linewidth]{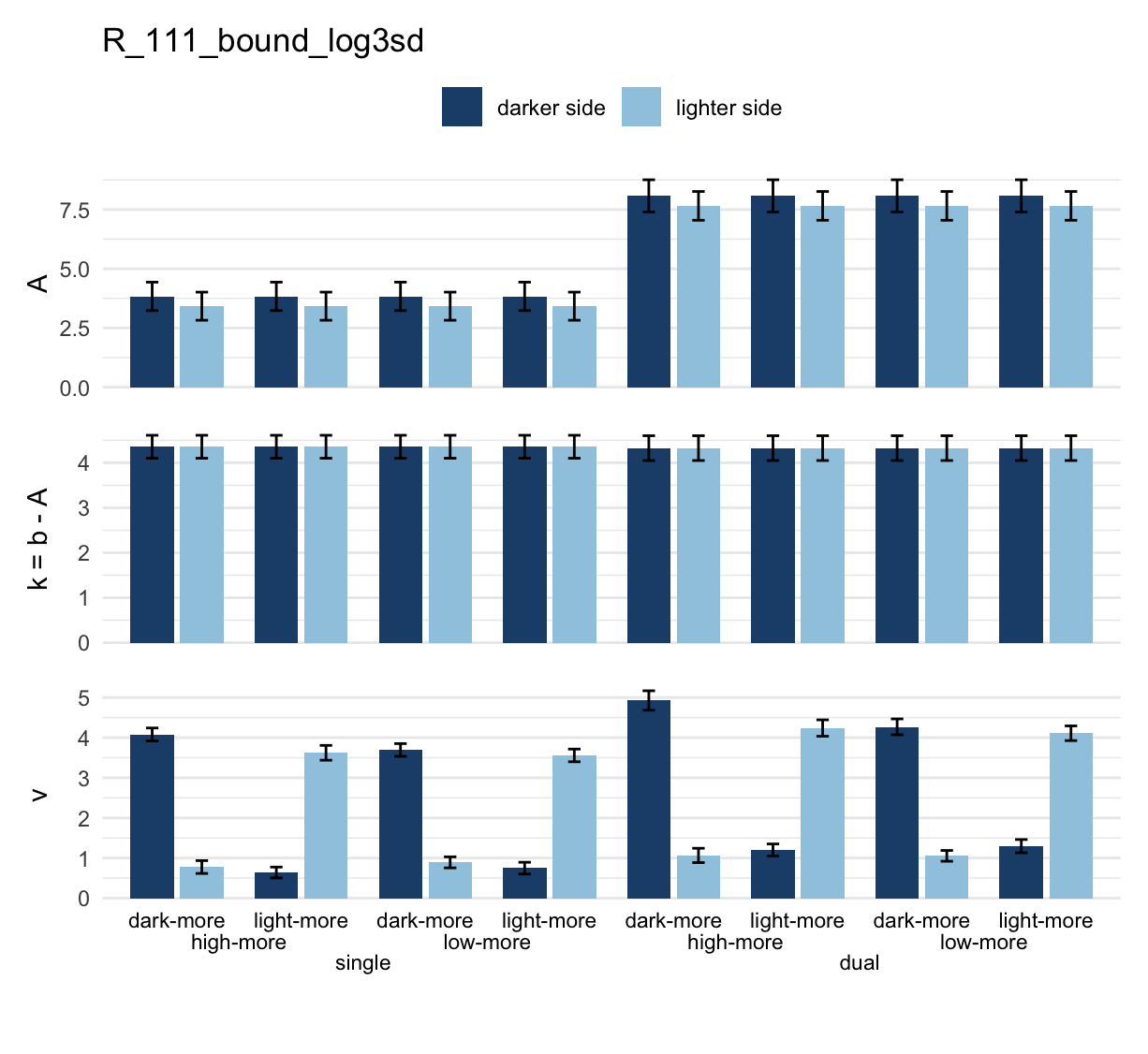}
    \caption{Mean parameter values for Experiment 1 final model.}
    \label{fig:lba_exp1_final_parameters}
\end{figure}

Figure~\ref{fig:lba_exp1_final_parameters} shows the mean parameter values of the final model fit to each participant in Experiment 1.

We fit linear mixed-effects regression models to predict parameter values ($A$, $k$, and $v$). For each factor over which these parameters were allowed to vary, we included the corresponding main effects and interactions as fixed effects, along with by-subject random intercepts and slopes. Table~\ref{tab:lba_exp1_final_parameters} show the results of the model.

\begin{table}[!h]
\small	

   \centering
   \caption{Experiment 1 final LBA model parameter analysis.}
   \begin{tabular}{lllccccc}
   \hline
        \textbf{Model} & \textbf{Predictor} & \textbf{$\beta$} & \textbf{SE} & \textbf{df} & \textbf{$t$} & \textbf{$p$}  \\
            \hline
$A$ & & Intercept  & 5.746  &  114.000 & 0.441 & 13.041 & $<.001$ \\
    & & Color  & 0.418  &  114.000 & 0.154 & 2.713 & $.008$ \\
    & & Task  & 4.239  &  114.000 & 0.881 & 4.810 & $<.001$ \\
    & & Color:Task  & 0.008  &  114.000 & 0.308 & 0.027 & $.979$ \\
\hline
    & Single & Intercept & 3.626 & 0.422 & 102 & 8.592 & $<.001$ \\
    & & Color & 0.414 & 0.844 & 102 & 0.491 & .625 \\
\hline
    & Dual & Intercept  & 7.865  &  0.455 & 126 & 17.269  & $<.001$ \\
    & & Color           & 0.422  &  0.911 & 126 & 0.464 & .644 \\
\hline
$k$ & & Intercept & 4.342  & 909.536 & 0.067 & 64.716 & $<.001$ \\
    & & Task	& -0.030 & 909.536 & 0.655 & -0.226 & 0.822 \\

\hline
$v$ & & Intercept  & 2.511  &  114.000 & 0.092 & 27.370 & $<.001$ \\
    & & Correctness  & 3.103  &  798.000 & 0.063 & 49.087 & $<.001$ \\
    & & Task  & 0.519  &  114.000 & 0.183 & 2.827 & $.006$ \\
    & & Lightness  & 0.168  &  798.000 & 0.063 & 2.654 & $.008$ \\
    & & Height  & 0.116  &  798.000 & 0.063 & 1.837 & $.067$ \\
    & & Correctness:Task  & 0.257  &  798.000 & 0.126 & 2.037 & $.042$ \\
    & & Correctness:Lightness  & 0.384  &  798.000 & 0.126 & 3.038 & $.002$ \\
    & & Task:Lightness  & -0.101  &  798.000 & 0.126 & -0.803 & $.422$ \\
    & & Correctness:Height  & 0.387  &  798.000 & 0.126 & 3.061 & $.002$ \\
    & & Task:Height  & 0.119  &  798.000 & 0.126 & 0.938 & $.349$ \\
    & & Lightness:Height  & 0.236  &  798.000 & 0.126 & 1.869 & $.062$ \\
    & & Correctness:Task:Lightness  & 0.455  &  798.000 & 0.253 & 1.800 & $.072$ \\
    & & Correctness:Task:Height  & 0.097  &  798.000 & 0.253 & 0.385 & $.700$ \\
    & & Correctness:Lightness:Height  & 0.376  &  798.000 & 0.253 & 1.489 & $.137$ \\
    & & Task:Lightness:Height  & 0.160  &  798.000 & 0.253 & 0.635 & $.526$ \\
    & & Correctness:Task:Lightness:Height  & 0.093  &  798.000 & 0.506 & 0.183 & $.855$ \\
\hline
    & Correct & Intercept  & 4.062  &  114.000 & 0.133 & 30.639 & $<.001$ \\
    & & Task  & 0.648  &  114.000 & 0.265 & 2.442 & $.016$ \\
    & & Lightness  & 0.360  &  342.000 & 0.042 & 8.645 & $<.001$ \\
    & & Height  & 0.310  &  342.000 & 0.042 & 7.439 & $<.001$ \\
    & & Task:Lightness  & 0.126  &  342.000 & 0.083 & 1.514 & $.131$ \\
    & & Task:Height  & 0.167  &  342.000 & 0.083 & 2.010 & $.045$ \\
    & & Lightness:Height  & 0.424  &  342.000 & 0.083 & 5.100 & $<.001$ \\
    & & Task:Lightness:Height  & 0.207  &  342.000 & 0.166 & 1.242 & $.215$ \\
\hline
    & Incorrect & Intercept  & 0.960  &  114.000 & 0.072 & 13.369 & $<.001$ \\
    & & Task  & 0.390  &  114.000 & 0.144 & 2.716 & $.008$ \\
    & & Lightness  & -0.024  &  342.000 & 0.095 & -0.255 & $.799$ \\
    & & Height  & -0.077  &  342.000 & 0.095 & -0.813 & $.417$ \\
    & & Task:Lightness  & -0.329  &  342.000 & 0.190 & -1.730 & $.085$ \\
    & & Task:Height  & 0.070  &  342.000 & 0.190 & 0.367 & $.714$ \\
    & & Lightness:Height  & 0.048  &  342.000 & 0.190 & 0.253 & $.801$ \\
    & & Task:Lightness:Height  & 0.114  &  342.000 & 0.380 & 0.300 & $.764$ \\
\hline
   \end{tabular}

   \label{tab:lba_exp1_final_parameters}

\end{table}

The results show that $A$, the starting evidence and its variance, is significantly higher in dual-task than in single-task. It is also significantly higher for the accumulator that corresponds to the darker side than the lighter side, suggesting higher average bias towards selecting the darker side even before looking at the stimulus. Meanwhile, $k$, the minimum amount of evidence needed before a response, is similar across tasks. This suggests that, combined with the difference in $A$, in the dual-task, the RT of each trial depends more on the starting evidence than in the single-task. This results in both faster and slower incorrect trials, which is what we actually observed in Experiment 1. 

Combined with $v$ being significantly higher in dual-task than in single-task, this suggests that participants are more \textit{rushed} in dual-task, prioritizing speed over accuracy than in single-task. They are overall faster at accumulating evidence, the pre-decisional bias ($A$) has more power over deciding how fast a trial is going to be completed.

\FloatBarrier

\subsubsection{Experiment 2 LBA models}

\vspace{0.5em}
\noindent\textbf{Free model after a priori assumptions}

\begin{figure}[H]
    \centering
    \includegraphics[width=0.7\linewidth]{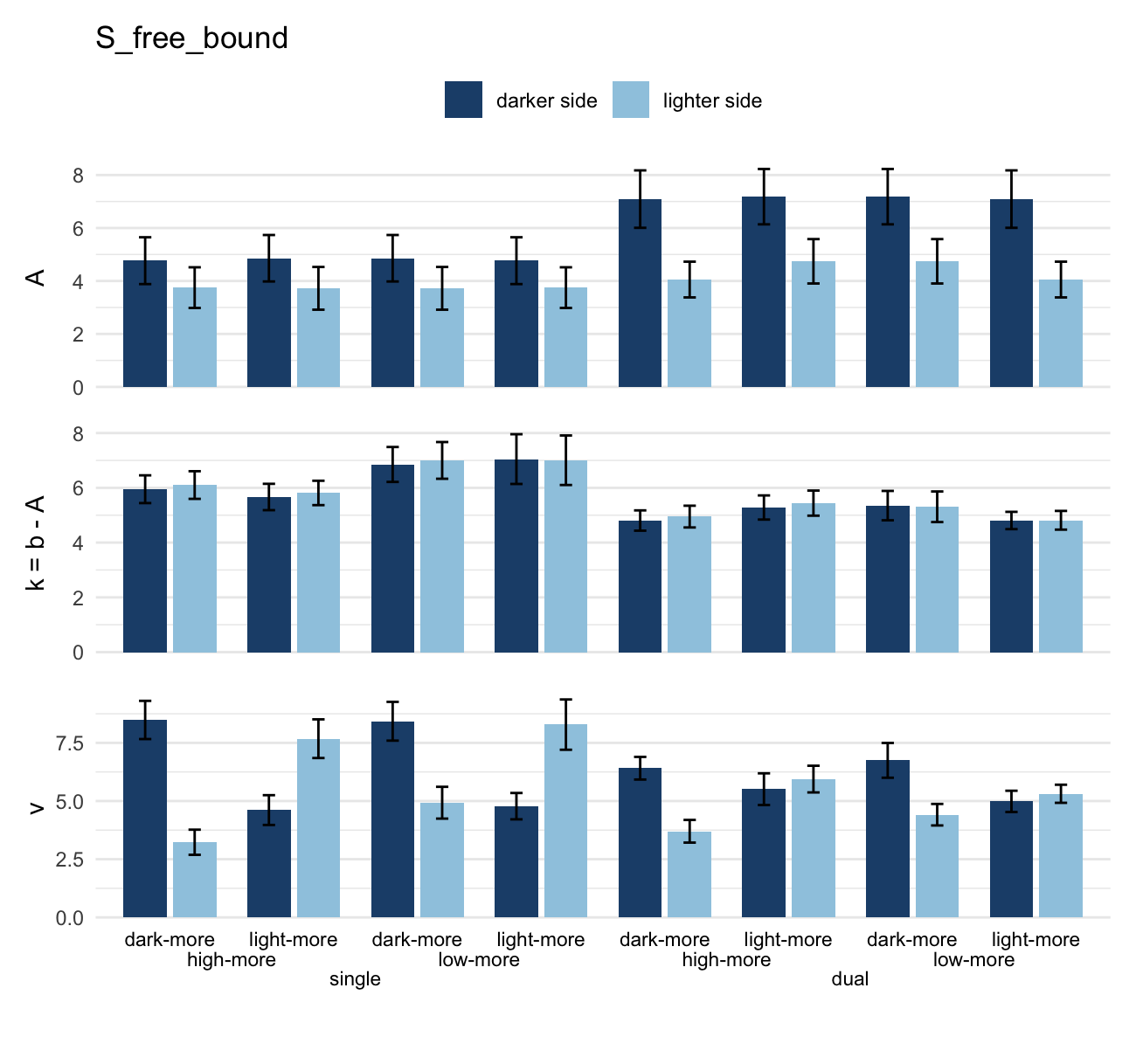}
    \caption{Parameters after fitting the free model to each participant's data. Bars show the mean of parameter values across participants. Error bars show standard error.}
    \label{fig:lba_exp2_free}
\end{figure}

Figure~\ref{fig:lba_exp2_free} shows the parameter values of maximally complex LBA model following the a priori assumptions, fit to each participant. We used the same a priori assumptions as Experiment 1, as described in Section \ref{sec:exp1_free}

This analysis allowed us to additionally 1) tie $A$ across \textit{legend-top color} and 2) tie $k$ across all conditions.

\vspace{0.5em}
\noindent\textbf{Candidate model evaluation}

There were a total of 8 candidate models. Table~\ref{tab:lba_exp2_candidates} shows the model specs of the 8 candidate models and their evaluations. Evaluations are based on AIC, BIC, and most importantly, whether the simulated data from the model can reproduce the behavioral pattern found in the original data.

The $\beta$ and $p$ in the table refers to the coefficient and the $p$-value associated with the three-way interaction between task, lightness mapping, and height mapping on Miss rate, which is what we found in the original data.

\begin{table}[!h]
\small
    \centering
    \caption{Candidate model specs and their evaluations based on our criteria. The $\beta$ and $p$ are based on LMER analysis on simulated data for Miss rate, and whether there is a 3-way interaction between task, lightness mapping, and height mapping.}
    \begin{tabular}{ccclllll}
    \hline
    \multicolumn{3}{c}{Parameter control}                                                            & \multicolumn{5}{c}{Evaluation}                                                 \\
    \cmidrule(lr){1-3} \cmidrule(lr){4-8}
    A: accumulator color & v: lightness mapping      & v: height mapping         & AIC & BIC & $\beta$ & $p$   & Recovery \\
    
    True                                     & True                      & True                      & 9791.309                & 13124.72                & -1.66   & 0.002 & yes      \\
    True                                     & True                      & False                     & 9772.417                & 11994.69                & -0.15   & 0.761 & no       \\
    True                                     & False                     & True                      & 10573.58                & 12738.69                & -0.22   & 0.646 & no       \\
    False                                    & True                      & True                      & 13209.08                & 10232.06                & -1.04   & 0.055 & no       \\
    True                                     & False & False & 15731.07                & 17397.77                & 0.14    & 0.776 & no       \\
    False                                    & False                     & True                      & 10734.32                & 12628.78                & 0.18    & 0.693 & no       \\
    False                                    & True                      & False                     & 34164.3                 & 36058.77                & -0.23   & 0.610 & no       \\
    False                                    & False                     & False                     & 15262.57                & 16615.77                & 0.21    & 0.654 & no     \\ 
    \hline
    \end{tabular}
    \label{tab:lba_exp2_candidates}
\end{table}

Since the three-way interaction could only be reproduced in the first candidate model, where $A$ varies by accumulator color and $v$ varies by lightness mapping, height mapping, and accumulator correctness. Therefore we select this model as our final LBA model.
Note that the candidate model where $A$ does not vary (4th row in the table) produced data that almost reached significance. We acknowledge that one might view this model as the better one, but we stuck to our predefined criteria.

\vspace{0.5em}
\noindent\textbf{Final model and Parameter Analysis}

We fit linear mixed-effects regression models to predict parameter values ($A$, $k$, and $v$). For each factor over which these parameters were allowed to vary, we included the corresponding main effects and interactions as fixed effects, along with by-subject random intercepts and slopes. Table~\ref{tab:lba_exp2_final_parameters} show the results of the model.

\begin{table}[H]
\small	
   \centering
   \caption{Experiment 2 final LBA model parameter analysis.}
   \begin{tabular}{llllccccc}
   \hline
        \textbf{Model} & & &  \textbf{Predictor} & \textbf{$\beta$} & \textbf{df} & \textbf{SE} & \textbf{$t$} & \textbf{$p$}  \\
            \hline
$A$ & & & Intercept  & 4.254  &  115.000 & 0.467 & 9.113 & $<.001$ \\
    & & & Task  & 1.175  &  115.000 & 0.934 & 1.258 & $.211$ \\
    & & & Color  & 1.488  &  115.000 & 0.202 & 7.368 & $<.001$ \\
    & & & Color:Task & 1.466  &  115.000 & 0.404 & 3.630 & $<.001$ \\
\hline
    & Single & & Intercept & 3.666  & 116 &   0.489  & 7.502 & $<.001$ \\
    & & & Color & 0.755 & 116 &   0.9774 & 0.773  &  0.441 \\
\hline
    & Dual & & Intercept &  4.841  & 114 & 0.466 & 10.395 & $<.001$ \\
    & & & Color & 2.221 & 114 & 0.931 &  2.385 & .019 \\
\hline 
$k$ & & & Intercept & 4.846 & 115 & 0.324 &	14.96	&$<.001$ \\
    & & & Task	&-0.889	&	115 & 0.648 & -1.372 & 0.173 \\
\hline
$v$ & & & Intercept & 4.842  &  115.000 & 0.286 & 16.944 & $<.001$ \\
    & & & Correctness & 2.171  &  805.000 & 0.123 & 17.637 & $<.001$ \\
    & & & Task & -0.834  &  115.000 & 0.572 & -1.459 & $.147$ \\
    & & & Lightness & -0.115  &  805.000 & 0.123 & -0.931 & $.352$ \\
    & & & Height & -0.022  &  805.000 & 0.123 & -0.178 & $.858$ \\
    & & & Correctness:Task & -1.565  &  805.000 & 0.246 & -6.356 & $<.001$ \\
    & & & Correctness:Lightness & 1.381  &  805.000 & 0.246 & 5.608 & $<.001$ \\
    & & & Task:Lightness & 0.117  &  805.000 & 0.246 & 0.476 & $.634$ \\
    & & & Correctness:Height & 0.824  &  805.000 & 0.246 & 3.345 & $<.001$ \\
    & & & Task:Height & -0.016  &  805.000 & 0.246 & -0.067 & $.947$ \\
    & & & Lightness:Height & -0.365  &  805.000 & 0.246 & -1.481 & $.139$ \\
    & & & Correctness:Task:Lightness & 0.291  &  805.000 & 0.492 & 0.591 & $.555$ \\
    & & & Correctness:Task:Height & -0.485  &  805.000 & 0.492 & -0.986 & $.325$ \\
    & & & Correctness:Lightness:Height & 1.543  &  805.000 & 0.492 & 3.133 & $.002$ \\
    & & & Task:Lightness:Height & 0.390  &  805.000 & 0.492 & 0.792 & $.428$ \\
    & & & Correctness:Task:Lightness:Height & -0.369  &  805.000 & 0.985 & -0.375 & $.708$ \\
\hline
    & Correct & & Intercept  & 5.928  &  115.000 & 0.351 & 16.898 & $<.001$ \\
    & & & Task  & -1.616  &  115.000 & 0.702 & -2.304 & $.023$ \\
    & & & Lightness  & 0.576  &  345.000 & 0.084 & 6.842 & $<.001$ \\
    & & & Height  & 0.390  &  345.000 & 0.084 & 4.632 & $<.001$ \\
    & & & Task:Lightness & 0.263  &  345.000 & 0.168 & 1.561 & $.119$ \\
    & & & Task:Height  & -0.259  &  345.000 & 0.168 & -1.539 & $.125$ \\
    & & & Lightness:Height  & 0.407  &  345.000 & 0.168 & 2.417 & $.016$ \\
    & & & Task:Lightness:Height  & 0.205  &  345.000 & 0.337 & 0.610 & $.542$ \\
\hline
    & & High-more & Intercept  & 6.123  &  115.000 & 0.363 & 16.874 & $<.001$ \\
    & & & Task  & -1.746  &  115.000 & 0.726 & -2.406 & $.018$ \\
    & & & Lightness  & 0.779  &  115.000 & 0.132 & 5.912 & $<.001$ \\
    & & & Task:Lightness  & 0.365  &  115.000 & 0.264 & 1.387 & $.168$ \\
\hline
    & & Low-more & Intercept  & 5.733  &  115.000 & 0.342 & 16.779 & $<.001$ \\
    & & & Task  & -1.487  &  115.000 & 0.683 & -2.176 & $.032$ \\
    & & & Lightness  & 0.372  &  115.000 & 0.126 & 2.944 & $.004$ \\
    & & & Task:Lightness  & 0.160  &  115.000 & 0.253 & 0.633 & $.528$ \\
\hline
    & Incorrect & & Intercept  & 3.757  &  115.000 & 0.254 & 14.762 & $<.001$ \\
    & & & Task  & -0.052  &  115.000 & 0.509 & -0.101 & $.919$ \\
    & & & Lightness  & -0.805  &  345.000 & 0.176 & -4.571 & $<.001$ \\
    & & & Height  & -0.434  &  345.000 & 0.176 & -2.463 & $.014$ \\
    & & & Task:Lightness  & -0.028  &  345.000 & 0.352 & -0.080 & $.936$ \\
    & & & Task:Height  & 0.226  &  345.000 & 0.352 & 0.643 & $.521$ \\
    & & & Lightness:Height  & -1.136  &  345.000 & 0.352 & -3.226 & $.001$ \\
    & & & Task:Lightness:Height  & 0.575  &  345.000 & 0.704 & 0.816 & $.415$ \\
\hline
    & & High-more & Intercept  & 3.540  &  115.000 & 0.257 & 13.780 & $<.001$ \\
    & & & Task  & 0.062  &  115.000 & 0.514 & 0.120 & $.905$ \\
    & & & Lightness  & -1.373  &  115.000 & 0.251 & -5.473 & $<.001$ \\
    & & & Task:Lightness  & 0.259  &  115.000 & 0.502 & 0.517 & $.606$ \\
\hline
    & & Low-more & Intercept  & 3.973  &  115.000 & 0.277 & 14.323 & $<.001$ \\
    & & & Task  & -0.165  &  115.000 & 0.555 & -0.297 & $.767$ \\
    & & & Lightness  & -0.237  &  115.000 & 0.263 & -0.899 & $.370$ \\
    & & & Task:Lightness  & -0.316  &  115.000 & 0.527 & -0.599 & $.550$ \\
\hline
   \end{tabular}
   \label{tab:lba_exp2_final_parameters}
\end{table}

\subsection{Experiment 1 Pruned Trials by Participant}
\label{sec:exp1_pruned_by_participant}

\includepdf[pages=-]{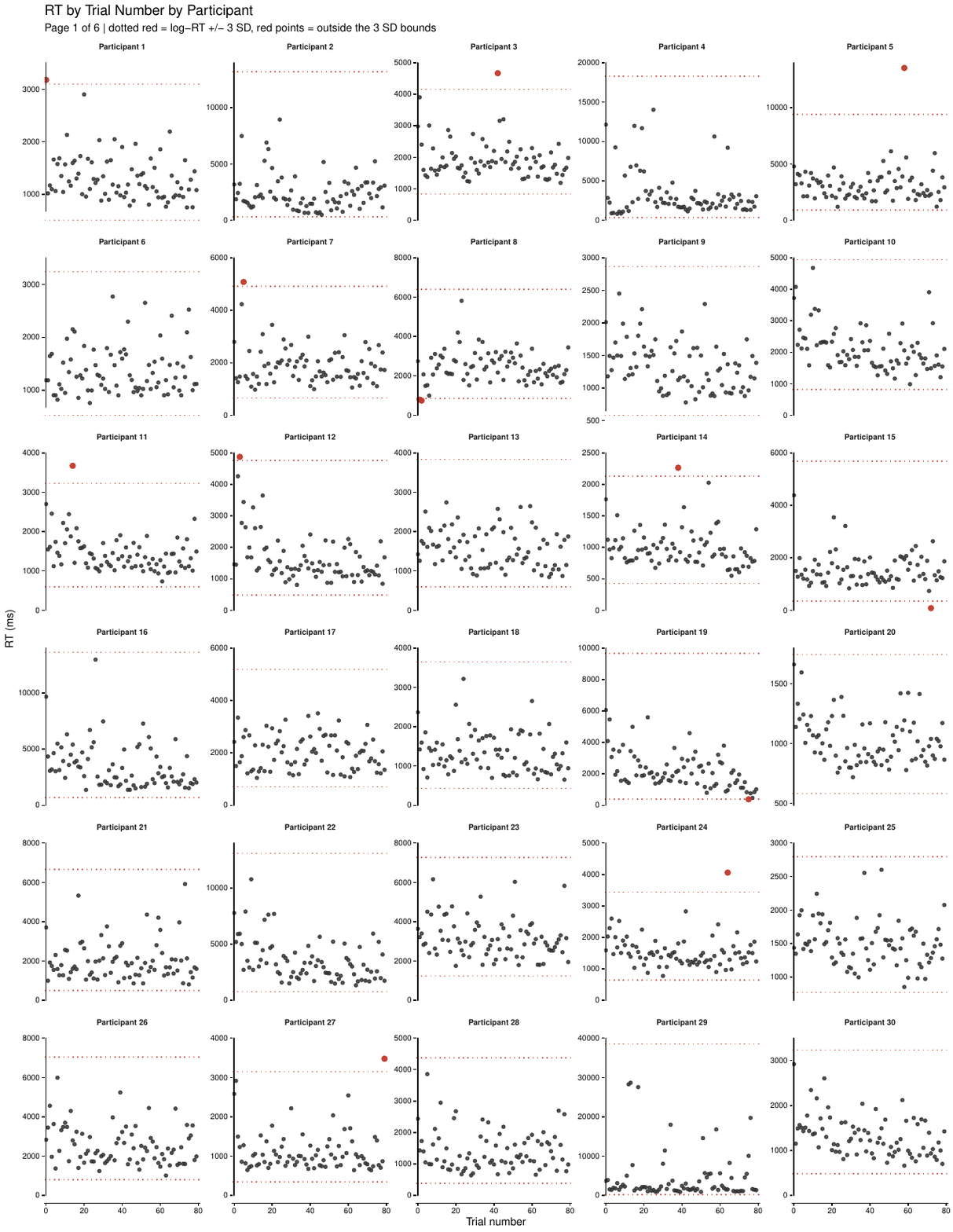}

\end{document}